\documentclass[10pt,journal,compsoc]{IEEEtran}
\ifCLASSOPTIONcompsoc
  \usepackage[nocompress]{cite}
\else
  \usepackage{cite}
\fi
\ifCLASSINFOpdf
\else
\fi
\usepackage{graphicx}                
\DeclareGraphicsExtensions{.pdf,.png,.jpg} 
\graphicspath{{./figures/}{./}} 
\makeatletter
\def\input@path{{./tables/}{./}}
\makeatother
\usepackage{microtype}                 
\PassOptionsToPackage{warn}{textcomp}  
\usepackage{textcomp}                  
\usepackage{mathptmx}                  
\usepackage{times}                     
\usepackage{cite}                      
\usepackage{tabu}                      
\usepackage{booktabs}                  
\usepackage{mathtools}
\usepackage{amsmath}
\usepackage{amssymb}
\usepackage{enumitem}
\usepackage{algorithmic}
\usepackage{grffile}
\usepackage{adjustbox}
\usepackage[outdir=./figures_eps/]{epstopdf}
\usepackage{algorithm}
\usepackage[table, dvipsnames]{xcolor}
\usepackage{tikz}
\usepackage{tikz-cd}
\usetikzlibrary{arrows.meta}
\usetikzlibrary{positioning}
\usepackage{wrapfig}
\usepackage{hyperref}

\DeclareMathAlphabet{\mathcal}{OMS}{cmsy}{m}{n}
\begin{document}
\renewcommand{\sectionautorefname}{Sec.}
\renewcommand{\subsectionautorefname}{Sec.}
\renewcommand{\figureautorefname}{Fig.}
\renewcommand{\equationautorefname}{Eq.}
\newcommand{\algorithmautorefname}{Alg.}
\renewcommand{\tableautorefname}{Tab.}
\newcommand{\mycaption}[1]{\vspace{-0ex}\caption{#1\vspace{-0ex}}}

\newcommand{\jules}[1]{\textcolor{black}{#1}}
\newcommand{\julien}[1]{\textcolor{black}{#1}}

\newcommand{\julienRevision}[1]{\textcolor{black}{#1}}
\newcommand{\julienMajorRevision}[1]{\textcolor{black}{#1}}
\newcommand{\julesRevision}[1]{\textcolor{black}{#1}}
\newcommand{\julesReplace}[2]{\textcolor{black}{#2}}
\newcommand{\julesReplaceMinor}[2]{\textcolor{black}{#2}}
\newcommand{\julienReplaceMinor}[2]{\textcolor{black}{#2}}
\newcommand{\julienReplace}[2]{\textcolor{black}{#1}}
\newcommand{\julesEditsOut}[1]{}

\title{A Progressive Approach to Scalar Field Topology}

\author{Jules~Vidal,
        Pierre~Guillou,
        and~Julien~Tierny
\IEEEcompsocitemizethanks{\IEEEcompsocthanksitem 
J. Vidal, P. Guillou, J. Tierny are with Sorbonne Université and CNRS.
\protect\\
E-mail: \{jules.vidal, pierre.guillou, julien.tierny\}@sorbonne-universite.fr
}
\thanks{Manuscript accepted Feb 17, 2021}}

\markboth{}%
{Shell \MakeLowercase{\textit{et al.}}: Bare Demo of IEEEtran.cls for Computer Society Journals}
\IEEEtitleabstractindextext{%
\begin{abstract}
This paper introduces progressive algorithms for the topological analysis of 
scalar data. 
Our approach is based on a hierarchical representation of the
input data and the fast identification of \emph{topologically invariant 
vertices}, \julienMajorRevision{which are vertices that have no impact on the 
topological description of the data and} for which we show that no computation 
is required as they are 
introduced in the hierarchy. This enables the definition of efficient 
coarse-to-fine topological algorithms, which leverage fast update mechanisms for 
ordinary vertices and avoid computation for the topologically invariant ones. We 
\julesReplace{instantiate}{demonstrate} 
our approach with two examples of topological algorithms (critical 
point extraction and persistence diagram computation), which generate 
\julesReplace{exploitable}{interpretable} outputs upon interruption requests and which progressively refine 
them otherwise.
Experiments on real-life datasets illustrate that our progressive strategy, in 
addition to the continuous visual feedback it provides, even improves run time 
\julesReplace{performances}{performance} with regard to non-progressive algorithms and we 
describe further accelerations with shared-memory parallelism.
We illustrate the 
utility of our approach in 
batch-mode and 
interactive 
setups, where it respectively enables 
the control of the 
execution time of complete topological pipelines as well as 
previews of the topological features found in a dataset, 
with progressive updates delivered within interactive times.
\end{abstract}

}

\maketitle

\IEEEdisplaynontitleabstractindextext

\IEEEpeerreviewmaketitle

\IEEEraisesectionheading{\section{Introduction}\label{sec:introduction}}

%
%
%
%

 

\newcommand{\domain}{\mathcal{M}}
\newcommand{\range}{\mathbb{R}}
\newcommand{\sublevelset}[1]{#1^{-1}_{-\infty}}
\newcommand{\Star}{St}
\newcommand{\Link}{Lk}
\newcommand{\simplex}{\sigma}
\newcommand{\face}{\tau}
\newcommand{\lowerlink}{\Link^{-}}
\newcommand{\upperlink}{\Link^{+}}
\newcommand{\Index}{\mathcal{I}}
\newcommand{\offset}{o}
\newcommand{\Natural}{\mathbb{N}}
\newcommand{\criticalSet}{\mathcal{C}}
\newcommand{\diagram}{\mathcal{D}}
\newcommand{\pointMetric}[1]{d_#1}
\newcommand{\wasserstein}[1]{W_#1}
\newcommand{\projection}{\Delta}
\newcommand{\hierarchy}{\mathcal{H}}
\newcommand{\decimation}{D}
\newcommand{\xDimD}{L_x^\decimation}
\newcommand{\yDimD}{L_y^\decimation}
\newcommand{\zDimD}{L_z^\decimation}
\newcommand{\xDim}{L_x}
\newcommand{\yDim}{L_y}
\newcommand{\zDim}{L_z}
\newcommand{\Grid}{\mathcal{G}}
\newcommand{\GridD}{\mathcal{G}^\decimation}
\newcommand{\x}{\phantom{x}}
\newcommand{\Mod}{\;\mathrm{mod}\;}
\newcommand{\NN}{\mathbb{N}}
\newcommand{\forwardIntegralLine}{\mathcal{L}^+}
\newcommand{\backwardIntegralLine}{\mathcal{L}^-}
\newcommand{\triangulationOp}{\phi}
\newcommand{\decimationOp}{\Pi}

\label{sec_intro}

\IEEEPARstart{T}{he} ever-increasing size and complexity of the datasets produced in 
engineering and sciences 
constitute 
a major
challenge for their
interpretation by human users.
To address these issues, advanced data analysis tools are designed to 
efficiently capture the main features of interest in large datasets, in order 
to support interactive visualization and analysis tasks. The tools developed 
in Topological Data Analysis (TDA) precisely serve this purpose. They form a 
family of techniques which 
focus on 
the generic, robust, and 
efficient extraction of structural features in data \cite{edelsbrunner09}. 
Over the last years, many 
data analysis and visualization methods have been built around these concepts
\cite{heine16}, with applications to a large spectrum of domains, 
including
astrophysics \cite{sousbie11, shivashankar2016felix},
biological imaging \cite{carr04, topoAngler, beiBrain18},
chemistry \cite{harshChemistry, chemistry_vis14, Malgorzata19}, 
fluid dynamics \cite{kasten_tvcg11},
material sciences \cite{gyulassy_vis07, gyulassy_vis15, soler_ldav19}, or
turbulent combustion \cite{laney_vis06, bremer_tvcg11, gyulassy_ev14}. In the 
case of scalar data, TDA algorithms rely on established 
topological data abstractions, such as 
contour trees 
\cite{boyell63, de1997trekking, tarasov98, carr00, smirnov17, gueunet_tpds19}, 
Reeb graphs 
\cite{reeb1946points, biasotti08, pascucci07}
or Morse-Smale complexes \cite{Defl15, gyulassy_vis08, 
robins_pami11, gyulassy_vis18}. 
In particular, the Persistence diagram \cite{edelsbrunner02} is a 
concise data representation, which visually summarizes the 
population of features of interest in a dataset, as a function of a measure of 
importance called \emph{topological persistence}. Its conciseness, combined 
with its stability, made it increasingly popular in machine learning
\cite{Carriere2017, ReininghausHBK15, rieck_topoInVis17} and in 
interactive data analysis,
where it quickly provides visual hints regarding the 
number and importance of the features in the data.

Although the core algorithms in TDA 
\julesReplace{admit}{have}
practicable asymptotic complexities 
(usually from linear to quadratic time), the construction of the above 
topological abstractions can still require significant computation times for 
real-life datasets. Thus, when they are integrated into larger interactive 
systems, TDA algorithms can become a serious time bottleneck.
This is a concern in data 
exploration scenarios, where users may need to wait between seconds and minutes 
to get a feedback when they adjust the parameters of the topological analysis.

In his seminal paper on response times of interactive systems,
Miller 
\cite{miller_response68}
studied the 
impact of
response time 
on the ability of users to maintain focus 
on a given task.
For response times below a second (\emph{continuity 
preserving latency}), the system 
appears fully responsive to user adjustments, allowing truly interactive 
sessions. 
For response times 
below
a few seconds (\emph{flow preserving 
latency}), users still manage to maintain their focus 
but the pauses in the exploration, due to the computation, challenge user 
interpretation
skills.
Above 
ten seconds
(\emph{attention preserving latency} 
threshold), users tend to disengage from the task to pursue other 
activities in parallel,
which is highly 
detrimental to the interpretation process.
Unfortunately, for real-life data, existing TDA algorithms correspond to the 
latter category of response times. 


To address the discontinuities in user experience implied by excessive 
computation times, the notion of \emph{progressive} data analysis has been 
explored by several authors in information visualization 
\cite{WilliamsM04,
fekete_arxiv16, 
zgraggen_tvcg17, jo_tvcg20}.
%
\begin{figure*}
  \includegraphics[width=\linewidth]{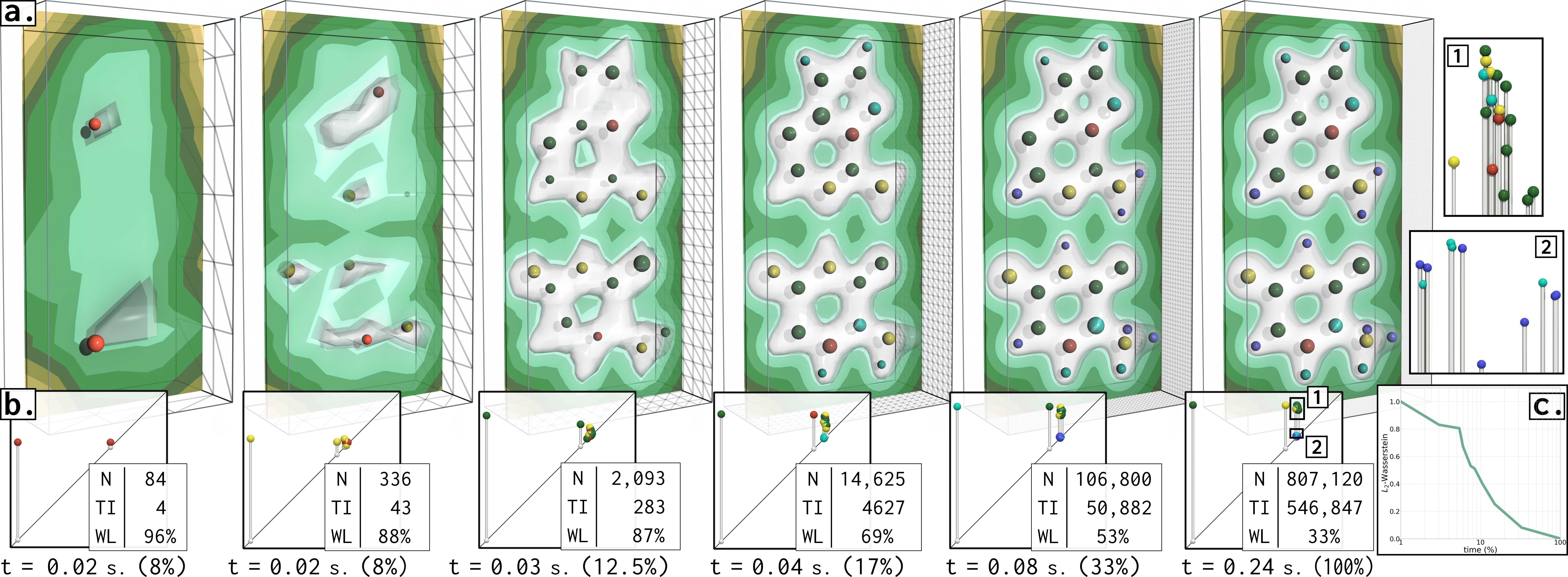}
  \mycaption{
  Progressive 
  persistence diagrams (saddle-maximum pairs) of 
the electron density 
of the adenine-thymine (AT) 
molecular system (an isosurface shows the two molecules), for 
a few steps of the progressive computation.
  Our coarse-to-fine approach efficiently refines the 
persistence diagram by progressing down a 
hierarchical
representation of the 
input (from left to right). Maxima (denoting the atoms) are shown in the 
domain (a) with spheres, scaled by topological persistence and colored by 
lifetime in the data hierarchy (from red to dark blue). 
\julienRevision{In this example, the}
persistence diagram (b) progressively captures the main features of the data.
As of 
\julien{8\%} of the computation 
time (leftmost), two persistent maxima are captured, denoting the presence of 
two molecules. 
As the computation progresses, atoms are progressively 
captured, heavier atoms first:  oxygens, then nitrogens and carbons, and 
finally hydrogens are respectively all captured as of 
\julien{12.5\%},
\julien{17\%} 
and \julien{33\%} of the computation time. At this point the diagram is 
complete and its 
accuracy is then improved 
until the final, exact 
result (rightmost). 
This qualitative progression is confirmed quantitatively by the empirical
convergence of the $L_2$-Wassserstein distance to the final output (c), which 
is monotonically decreasing: more computation time indeed yields more accuracy.
Our algorithms leverage efficient update mechanisms and
\emph{topologically invariant vertices} (\emph{TI}), which can be quickly
identified 
and for which we show that no computation is 
required, thus drastically reducing the workload (\emph{WL}) of the algorithm 
with time.
Overall, 
our progressive approach efficiently computes the persistence diagram of the 
data,
while continuously providing relevant visual feedback.
        }
        \label{fig:teaser}
\end{figure*}
In this context, a progressive algorithm is a technique capable of providing an 
\julesReplace{exploitable}{interpretable} output upon 
interruption requests, and of refining it otherwise,  progressively converging 
towards the final solution.
In an interactive setup, progressive algorithms can improve user experience in 
two ways. First, users can let such algorithms refine progressively their 
outputs, while receiving continuously visual feedbacks, and stop them when the 
outputs are deemed satisfactory. Second, users can also define a priori an 
upper 
limit on the computation time, after which the computation is interrupted. 
This enables to design interactive systems with guaranteed response time.
This is particularly useful for algorithms whose actual execution times are 
difficult to anticipate, such as in TDA, where many popular algorithms, 
although with known time complexity, may 
\julesReplace{admit}{have} an output-sensitive computation time in practice.
Progressive algorithms can also be beneficial to non-interactive setups,
such 
as batch-mode computations on high performance computers, where the allocation 
of computing \julienRevision{resources} often needs to be finely controlled.
In this context, 
\julesEditsOut{progressivity can help optimize scheduling, as progressive tasks 
can be put on pause, and still provide an exploitable result which can be later 
refined upon computation resume.}
\julienReplace{progressivity enables the assignment of precise}{Progressivity 
also 
enables the assignment of a}{} computation time budget to 
data analysis programs.
\julienReplace{}{\julesRevision{ since they enable the assignment of a 
computation time budget to data analysis programs}.}
This is relevant for time-critical applications such as numerical 
simulation in support of urgent decision making in catastrophe management
(wildfires, floods, outbreaks, etc). 
This is also relevant to more general applications, towards the
control of the power consumption of high performance computers, which 
becomes an increasingly important societal issue.

In this work, we introduce, to our knowledge, the first progressive algorithms 
for the topological analysis of scalar data. Our overall approach is based on 
the key idea that \emph{critical points} (which 
correspond to topological events in data, \autoref{sec_criticalPoints}) 
correspond to singular events, which usually have a reasonably low probability 
of appearance\julesReplace{, and that most of the points of a dataset}{. In
particular, most of the points of a dataset are regular points 
\julienMajorRevision{in practice}, and} do not 
imply topological events\julesReplace{ in general}{}. For those, fast update 
mechanisms can be derived, allowing for efficient progressive algorithms.
Our overall approach is based on a 
hierarchical
representation of 
the data (\autoref{sec_progressiveData}), where \emph{topologically invariant 
vertices} (vertices of the input which do not produce topological 
events upon insertion, \julienMajorRevision{and which, therefore, have no 
impact on the topological representation of the data,} 
\autoref{sec_topologicalInvariants}) 
can be identified and processed very efficiently. This enables the definition 
of coarse-to-fine 
topological algorithms, which provide an exact output for each level of the 
data
hierarchy, and which quickly update their outputs between 
consecutive hierarchy levels.
\julienRevision{This strategy is motivated by two key practical observations: 
\emph{(i)}
the main features of interest of a dataset often appear 
early in the data hierarchy
in practice 
and \emph{(ii)} critical points represent a 
small portion of the input vertices, allowing for fast update mechanisms 
leveraging the inherent regularity of the data (when present). These two 
practical observations are quantitatively evaluated in our experiments 
(\autoref{sec_results}).}
We \julesReplace{instantiate}{demonstrate} our progressive strategy with two 
examples of topological analysis algorithms: critical point extraction 
(\autoref{sec_progressiveCriticalPoints}) and persistence diagram computation 
(\julienRevision{for 
extremum-saddle pairs}, \autoref{sec_progressivePersistenceDiagram}). 
In  both cases, the inherent regularity of the data 
is
leveraged to derive efficient progressive algorithms, which continuously 
provide visual feedback, and which compute the final result 
even faster in 
practice than 
existing non-progressive methods. We also present simple parallelizations of  
our algorithms,
to further improve 
\julesReplace{performances}{performance}.
Experiments on real-life datasets validate the relevance of our progressive 
representations, 
both at a qualitative 
and quantitative level. 
\julien{We illustrate the practical utility of our approach, both for 
batch-mode and interactive setups, where it respectively enables the
control of the run time of a TDA pipeline, as well as  
\julienRevision{progressive}
previews \julienRevision{of the topological features found 
in a dataset, continuously}
updated within interactive times.}
%


\subsection{Related Work}
\label{sec_relatedWork}

Many approaches based on topological methods have been documented over the last 
two decades. We refer the reader to the survey by Heine~et~al.~\cite{heine16} 
for a comprehensive overview. 
In the following, we focus on algorithms for 
constructing topological data abstractions, which \julesReplace{is}{are} the most related to our 
work.

While Morse theory has originally been developed in the smooth setting 
\cite{milnor63}, many of its concepts can be translated to discretized data, in 
particular in the form of piecewise-linear (PL) scalar fields defined on PL 
manifolds. 
Banchoff~\cite{banchoff70} introduced a formalism for the 
combinatorial characterization of 
the \emph{critical points} (\autoref{sec_criticalPoints}) of an input PL scalar 
field. These points correspond to 
locations where the sub-level sets of the 
function change their topology. They correspond to notable events in the data. 
\julesReplace{For instance,}{In practice,} extrema are often associated \julesEditsOut{in practice }with features of 
interest. In presence of noise however, many critical points can be reported by 
this characterization, which motivates the introduction of an importance 
measure on critical points, to distinguish noise artifacts from salient 
features. 

Topological persistence \cite{edelsbrunner02, edelsbrunner09} 
\julesReplaceMinor{established itself}{has been established} as a reference measure to assess the importance of critical 
points. It can be directly read from the \emph{Persistence diagram} 
(\autoref{sec_persistenceDiagram}) which plots \julesEditsOut{the }topological features of the 
data according to their \emph{birth} and \emph{death}, both of which exactly 
coincide with critical points. Thus, the critical points of the input data are 
arranged in the diagram in pairs. The Persistence diagram can be computed 
generically by matrix reduction operations \cite{edelsbrunner02, 
edelsbrunner09}.
The pairs of critical points in the diagram 
which involve extrema,\julesEditsOut{ which are} often associated to features of interest in 
applications, can be 
computed more efficiently, with a Union-Find data 
structure \cite{edelsbrunner09, cormen}, or equivalently, they can be read 
directly 
from the merge tree (presented further down). For the special case of 
point cloud data, the 
topology of the underlying manifold sampled by the point cloud can be inferred 
\cite{ChazalO08}
by considering the persistence diagram of the Vietoris-Rips filtration, for 
which tailored algorithms have been developped \cite{ripser}.

Although persistence diagrams are stable \cite{CohenSteinerEH05}
\julienReplace{(i.e. a small perturbation to the input data will only yield a 
small perturbation of its persistence diagram), }{,
\julesRevision{meaning that similar scalar fields exhibit similar diagrams,}} 
their 
discriminative power may be insufficient for some applications. 
This motivates the design of more discriminative topological 
abstractions, such as \julesReplaceMinor{the }{}merge and contour trees, which \julesEditsOut{respectively }track the 
connected components of sub-level sets and level sets. 
\julienMajorRevision{Intuitively, these trees indicate how level sets 
connect and disconnect
when passing critical points.}
The first algorithms for 
computing these tree structures focused on the 2D \cite{boyell63, 
de1997trekking} and 3D \cite{tarasov98} cases. 
In their 
seminal paper, Carr~et~al.~\cite{carr00} introduced an efficient algorithm, 
with optimal time complexity, for computing the contour tree in all dimensions.
Recently, several algorithms have been documented to compute this 
structure in parallel \cite{
PascucciC03,
MaadasamyDN12,
MorozovW14,
AcharyaN15,
CarrWSA16, 
gueunet_ldav17,
smirnov17, 
gueunet_tpds19}.
If the input domain is not simply connected
\julienMajorRevision{(intuitively, if it contains handles)}, the Reeb graph 
\cite{reeb1946points} 
needs to 
be 
considered instead of the contour tree to correctly  track connected 
components of 
level sets, which involves more sophisticated methods 
\julienMajorRevision{(as the Reeb graph may now contain loops)}. 
The first Reeb graph computation algorithms 
were based on a 
slicing strategy \cite{ShinagawaKK91, BiasottiFS00, WoodHDS04}, 
later 
improved by solely slicing along critical contours \cite{PataneSF08, 
DoraiswamyN12}. Several techniques focused on practical \julesReplace{performances}{performance} 
\cite{pascucci07, tierny_vis09}, while algorithms with optimal time complexity 
have been introduced, first in 2D \cite{ColeMcLaughlinEHNP03}, later
in arbitrary 
dimension \cite{Parsa12}, and then parallelized 
\cite{GueunetFJT19}. Recently, 
efficient algorithms have been investigated  for the computation of the 
generalization of the Reeb graph to 
multivariate functions, called the Reeb space \cite{
EdelsbrunnerHP08, 
CarrD14, tierny_vis16}.

The Morse-Smale complex is another typical topological abstraction for scalar 
data \cite{Defl15}. It decomposes the input domain into cells which \julesReplace{admit}{have}
identical gradient integration extremities. 
\julienMajorRevision{Intuitively, it 
segments the data into regions, bounded by gradient flow separatrices, where 
the gradient shows a homogeneous behaviour.}
While the initial algorithms for 
its computation were developed in the PL setting 
\cite{
EdelsbrunnerHZ01,
EdelsbrunnerHNP03
}, modern alternatives \cite{gyulassy_vis08, robins_pami11} 
are based on Discrete Morse theory \cite{forman98} and parallel algorithms have 
been documented \cite{ShivashankarN12, gyulassy_vis18}.

To our knowledge, no algorithm has been described so far for the 
\emph{progressive} computation of the above structures. In this work, we 
introduce the first progressive algorithms for the computation of topological 
abstractions, namely critical points (\autoref{sec_progressiveCriticalPoints}) 
and persistence diagrams 
(\julienRevision{for 
extremum-saddle pairs, }\autoref{sec_progressivePersistenceDiagram}).
Our approach is based on a 
hierarchical
representation of the data. Multiresolution hierarchies have been 
considered 
before, for the Reeb graph \cite{HilagaSKK01}, the contour tree 
\cite{pascucci_mr04} and the Morse-Smale 
complex\julienRevision{~\cite{BremerEHP03, gunther2012, IuricichF17}}, but the 
hierarchical
aspect dealt with the \emph{output} data structure, while the 
input was processed without multiresolution, with existing algorithms 
\cite{BiasottiFS00, carr00, EdelsbrunnerHZ01}.
In contrast, in our work, the \emph{input} data is represented \julesReplaceMinor{in}{as} a 
multiresolution hierarchy and the output is efficiently, progressively 
updated \julienRevision{in a coarse-to-fine manner,} by iterating 
through the hierarchy levels.

\julienRevision{Our progressive scheme relies on a hierarchical representation 
of 
the input data. In the visualization community, many types of hierarchies have 
been defined to encode 
and extract visual representations from volumetric data at different levels of 
details \cite{GregorskiDLPJ02, weiss_vis09, weiss_sgp09, PascucciB00, 
LewinerVLM04, GerstnerP00}. 
\julienMajorRevision{For example, Gerstner and Pajarola \cite{GerstnerP00} 
introduce a method for the 
robust extraction of isosurfaces in 
multiresolution volume representations. For this, their algorithm extracts
the critical points of the input scalar field, for each level of their
hierarchical scheme. However, they use for this the standard, non-progressive, 
procedure 
\cite{banchoff70}. In contrast, our approach extracts the critical points for 
all of our hierarchy levels \emph{progressively}, i.e. without recomputing from 
scratch 
critical points at each new hierarchy level, but instead
by efficiently and minimally 
updating the information already computed at the previous levels.}
\julienReplace{Generally, in}{In} our work, we focus on a 
specific scheme based on the so-called
\emph{red} subdivision \cite{freudenthal42, bank83, loop87, 
bey95, zhang95} applied to regular grids \cite{kuhn60, bey98}, in particular to 
investigate progressive and efficient \emph{coarse-to-fine} computations,
in contrast to 
the traditional fine-to-coarse hierachical approaches found in the 
visualization literature.}


%

The approaches which are the most related to our work are probably the 
streaming algorithms 
for computing  the Reeb graph \cite{pascucci07}
and the merge tree\cite{bremer_tvcg11}. 
These algorithms are capable of computing 
their output in a streaming fashion: the simplices of the input domain can be 
processed in arbitrary order and these algorithms maintain and iteratively 
complete their output data structure. However, while they can be interrupted, 
these algorithms are not, strictly speaking, \emph{progressive}: upon 
interruption, they do not provide \julesReplace{exploitable}{interpretable} but \emph{partial} results, which 
are very far in practice from the final result. 
For instance, the streaming Reeb graph \cite{pascucci07} can typically count at 
a given time
a very large number of loops (which will be iteratively filled
as the algorithm progresses). In
contrast, our 
coarse-to-fine algorithms provide \julesReplace{exploitable}{interpretable} 
results upon interruption, which are visually similar to the exact, final 
outputs and which empirically quickly converge towards them.

\subsection{Contributions}
\label{sec_contributions}

This paper makes the following new contributions:

\begin{enumerate}[leftmargin=1em]
 \item \emph{A progressive \julien{data representation} 
(\autoref{sec_progressiveData})} We present an approach for the progressive 
topological analysis of scalar data, to generate \julesReplace{exploitable}{interpretable} outputs 
upon interruption requests. 
Our approach \julienRevision{relies} on a hierarchical
representation of the input \julienRevision{data (derived from established 
triangulation subdivision schemes \cite{freudenthal42, kuhn60, bank83, loop87, 
bey95, zhang95, bey98})} and the fast identification of \julienRevision{the new 
notion of}
\emph{topologically invariant vertices}, 
for which we show that no computation is required
as they are introduced in the hierarchy.

 
 \item \emph{A progressive algorithm for critical point extraction 
(\autoref{sec_progressiveCriticalPoints})} 
We introduce a progressive algorithm 
for critical point extraction.
As it progresses down the data hierarchy, our algorithm 
leverages efficient update mechanisms for ordinary vertices and avoids 
computation for the topologically invariant ones. This enables a
progressive output refinement, which \julesReplace{even results in}{results in even} faster overall 
computations
than non-progressive methods.
We 
also 
introduce
a fast heuristic to evaluate 
the lifetime of critical points in the data hierarchy.
%

\item \emph{A progressive algorithm for persistence diagram computation
(\autoref{sec_progressivePersistenceDiagram})} We introduce a progressive 
algorithm for the computation of persistence diagrams of 
extremum-saddle pairs, 
built on top of the above contributions. In practice, our algorithm 
tends to capture the main features of the data first, and then progressively 
improves its accuracy. This is confirmed quantitatively by the empirical 
convergence of the Wasserstein distance to the final output, which is 
monotonically decreasing (more computation time indeed yields more accuracy). 
Our approach enables a 
continuous visual feedback,
while being in practice even faster overall than non-progressive methods.

\item \emph{A reference implementation}
We provide a reference C++ implementation of our algorithms 
\julienRevision{(publicly available at:
\href{https://github.com/julesvidal/progressive-scalar-topology}{
https://github.com/julesvidal/progressive-scalar-topology})} that can 
be used to replicate our results\julienRevision{, and} 
for future 
benchmarks.

\end{enumerate}


\section{Preliminaries}
\label{sec_preliminaries}
This section briefly presents the technical background of our work. We refer 
the reader to the \julesReplace{reference }{}textbook by Edelsbrunner and Harer 
\cite{edelsbrunner09} for a detailed introduction to computational topology.

\subsection{Input Data}
\label{sec_plScalarField}
The input is modeled as a piecewise linear (PL) scalar 
field $f : \domain 
\rightarrow \range$ defined on a PL $d$-manifold $\domain$, with $d$ equals 2 
or 3 in our applications. 
The scalar values are given at the vertices of 
$\domain$ and are linearly interpolated 
on the other
simplices \julienMajorRevision{(with barycentric coordinates)}.
$f$ is assumed to be injective on the vertices 
of $\domain$ \julienMajorRevision{(i.e. each vertex has a distinct $f$ value)}. 
This is enforced in practice with a symbolic 
perturbation inspired by Simulation of Simplicity \cite{edelsbrunner90}. 
Specific requirements
on the structure of the triangulation $\domain$ are discussed 
in Secs. \ref{sec_multiRes} and \ref{sec_limitations}.

\subsection{Critical Points}
\label{sec_criticalPoints}
\julesReplace{The topological}{Topological} features of $f$ can be tracked with the notion of 
\emph{sub-level} set, noted
$\sublevelset{f}(w) = \{ p \in \domain ~ | ~ f(p) < w\}$. 
\julienMajorRevision{It is simply the subset of the data below a certain 
threshold $w$.}
In particular,
the topology of these sub-level sets (in 3D their connected components,
cycles and voids) can only change at specific locations, named the
\emph{critical points} of $f$ \cite{milnor63}. In the PL setting, Banchoff
\cite{banchoff70} introduced a local characterization of critical
points, defined as follows.

\julesReplaceMinor{}{A \textit{face} $\tau$ of a simplex $\sigma \in \domain$ is a simplex of 
$\domain$ that is defined by a non-empty, strict subset of the 
\julienReplaceMinor{points}{vertices} of 
$\sigma$.
We call $\sigma$ a \textit{co-face} of $\tau$ and we note
$\tau<\sigma$.}
The \emph{star} of a vertex $v \in \domain$, noted $\Star(v)$,  is
the set of its co-faces:
$\Star(v) = \{ \simplex \in \domain ~|~ v < \sigma \}$. 
\julienMajorRevision{This can be viewed as a small, combinatorial, neighborhood 
around $v$.}
The 
\emph{link} of $v$,
noted $\Link(v)$, is the set of the faces $\face$ of the simplices $\simplex$
of $\Star(v)$ with empty intersection with $v$:
$\Link(v) = \{ \face \in \domain ~ | ~ \face < \simplex, ~
\simplex\in \Star(v), ~ \face \cap v = \emptyset\}$. 
\julienMajorRevision{This can be viewed as the \emph{boundary} of a small, 
combinatorial, neighborhood around $v$.}
The \emph{lower link} of $v$, noted $\lowerlink(v)$, is given by the 
set of simplices of $\Link(v)$ which only contain vertices \emph{lower} than 
$v$:
$\lowerlink(v) = \{ \simplex \in \Link(v) ~ | ~ \forall v' \in \sigma, ~ f(v')
< f(v)\}$. The upper link is defined symmetrically: $\upperlink(v) = \{ 
\simplex \in \Link(v) ~ | ~
\forall v' \in \sigma, ~ f(v') > f(v)\}$.
A vertex $v$ is \emph{regular}  if
both $\lowerlink(v)$ and $\upperlink(v)$ are simply connected. For
such vertices, the sub-level sets 
\julienReplace{enter the neighborhood of $v$, $\Star(v)$, through the 
lower part of the neighborhood boundary, $\lowerlink(v)$, and exit through its 
upper part, $\upperlink(v)$, \emph{without} changing their topology.}{
do not change their topology as they span
$\Star(v)$.} Otherwise, $v$ is 
a \emph{critical point}.
These can be classified with regard to their
\emph{index} $\Index(v)$\julienMajorRevision{, which intuitively corresponds 
to the number of independent directions of decreasing $f$ values around $v$}.
It is equal to $0$ for local minima
($\lowerlink(v) = \emptyset$), to $d$ for local maxima
($\upperlink(v) = \emptyset$) and otherwise to $i$ for
$i$-saddles ($0 < i < d$).

Adjacency 
relations between critical points can be captured with the notion of 
\emph{integral line}. Given a vertex $v$, 
its \emph{forward} integral 
line, noted $\forwardIntegralLine(v)$, is a path along the edges of 
$\domain$, 
initiated in $v$, such that each edge of $\forwardIntegralLine(v)$ connects a 
vertex $v'$ to its highest neighbor $v''$. 
Then, forward integral lines are 
guaranteed to terminate in local maxima of $f$. 
When encountering a saddle $s$, we 
say that an integral line \emph{forks}: it yields one new integral line per 
connected component of $\upperlink(s)$. 
Note that several integral lines can \emph{merge} (and possibly fork later). A 
\emph{backward} integral line, noted $\backwardIntegralLine(v)$ is defined 
symmetrically (i.e. integrating downwards towards minima).
Critical points play a central role in TDA
as
they often correspond to features of interest in various applications: centers 
of vortices in fluid dynamics \cite{kasten_tvcg11}, atoms in chemistry
\cite{harshChemistry, chemistry_vis14, Malgorzata19} or clusters of galaxies in
astrophysics \cite{sousbie11, shivashankar2016felix}.

\begin{figure}
 \centering
 \includegraphics[width=0.95\linewidth]{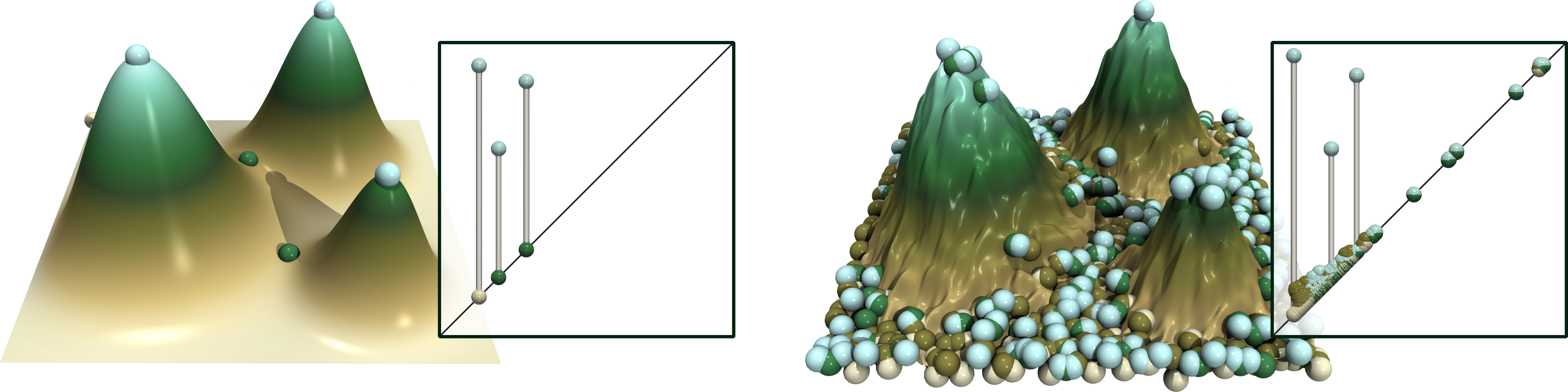}
 \mycaption{Persistence diagrams of a clean (left) and noisy (right)
 scalar 
field 
 (light brown spheres: minima, cyan: maxima, others: saddles). The main three 
hills 
are clearly apparent in the diagrams (high persistence pairs), whereas 
small pairs near the diagonal indicate noisy features.}
 \label{fig_pd}
\end{figure}

\subsection{Persistence Diagrams}
\label{sec_persistenceDiagram}
Several importance 
measures for critical points have been studied \cite{carr04},  
including \emph{topological 
persistence} \cite{edelsbrunner02}, which is tightly coupled to the notion of 
Persistence diagram \cite{edelsbrunner09}, which we briefly summarize here.
\julienMajorRevision{In practical applications, features of interest are 
often characterized by the extrema of the field. Thus, in the following, we 
will first focus our description on local minima, and then discuss 
generalizations. The importance of a local minimum can be assessed with its 
\emph{persistence}, which describes the lifetime of the topological feature 
(i.e. the connected component) it created in $\sublevelset{f}(w)$.}
%
\julesReplace{In particular, a}{A}s $w$ increases, new connected components of 
$\sublevelset{f}(w)$ are created at the minima of $f$. The Elder rule 
\cite{edelsbrunner09} indicates that if two connected 
components, created at the minima $m_0$ and $m_1$ with $f(m_0) < f(m_1)$, meet 
at a given $1$-saddle $s$, the \emph{youngest} of the two components (the 
one created at $m_1$) \emph{dies} in favor of the \emph{oldest} one (created at 
$m_0$). In this case, a \emph{persistence pair} $(m_1, s)$ is created and 
its 
\emph{topological persistence} $p$ is given by $p(m_1, s) = f(s) - f(m_1)$. 
All \julesReplaceMinor{the }{}local minima
can be 
unambiguously 
paired following this strategy, while the
global minimum is usually paired, by convention, with the global maximum.
\julesReplace{}{\julienMajorRevision{Symmetrically}, 
persistence assesses the importance of a 
local maximum paired with a \julienReplace{$(d-1)$}{$2$}-saddle,
based on the 
lifetime of the topological feature it destroyed in 
$\sublevelset{f}(w)$.
}

\julesRevision{Generally, as one continuously increase\julesRevision{s} an isovalue $w$, 
  topological structures in $\sublevelset{f}(w)$ (connected components, cycles, 
  voids) are created and destroyed at 
  critical points.
\julienReplace{Thus}{As such}, each topological feature is 
characterized by a pair of critical points, \julienReplace{which 
indicate its birth 
and death, and whose difference in function values indicates its lifespan 
in the data, its \emph{persistence}.}{The topological persistence
of these features denote their lifespan
in term of the $f$ values.}}
\julesReplace{The symmetric reasoning can be applied in 3D to characterize,
with $2$-saddle/maximum pairs, the life time of the voids of
$\sublevelset{f}(w)$, while the $1$-saddle/$2$-saddle pairs characterize its
independent cycles.}{As described above, the persistence of connected
components of $\sublevelset{f}(w)$ is encoded \julienReplace{with}{in}
minimum/$1$-saddle pairs. In
3D, $2$-saddle/maximum pairs characterize the life time of the voids of
$\sublevelset{f}(w)$, while \julesReplaceMinor{the }{}$1$-saddle/$2$-saddle pairs characterize its
independent cycles.}
\julienReplace{As mentioned above}{In practical applications}, features of 
interest are often characterized by the 
extrema of the field. Thus, in the following, when considering 
persistence diagrams, we will focus on minimum/$1$-saddle pairs 
and $(d-1)$-saddle/maximum pairs.

Persistence pairs are usually 
visualized with the \emph{Persistence diagram} $\diagram(f)$ 
\cite{edelsbrunner09}, which embeds each pair $(c, c')$, with $f(c) < f(c')$, 
as a point in the 2D plane, at location $\big(f(c), f(c')\big)$. There, the 
pair 
persistence
can be 
visualized as the height of the point to the diagonal. 
\julienMajorRevision{In other words, in the persistence diagram, each 
topological feature 
of $\sublevelset{f}(w)$ (connected component, cycle, void) can be 
readily visualized as a bar (\autoref{fig_pd}), whose height to the diagonal 
denotes its importance in the data.}
Features with a high persistence stand out, away from the diagonal, 
while noisy features are typically located in its vicinity. 
The conciseness, stability \cite{edelsbrunner02} and expressiveness of this 
diagram made it a popular tool 
for data summarization tasks. 
As shown in \autoref{fig_pd},
it provides visual hints about the number, ranges and salience 
of the features of interest.

\julesEditsOut{To evaluate quantitatively  the relevance of the progressive outputs
continuously provided by our algorithms, we measure their distance 
to the final, exact result with the \emph{Wasserstein distance}, an established 
practical metric 
inspired by optimal 
transport \cite{Kantorovich, monge81}.
Given two diagrams $\diagram(f)$ and
$\diagram(g)$, a pointwise distance 
 $\pointMetric{q}$, inspired from the $L^p$ norm, can be introduced 
in the 2D birth/death space
between 
  two points $a = (x_a, y_a) \in \diagram(f)$ and 
$b = (x_b, y_b) \in \diagram(g)$, with $q > 0$, as: 
\vspace{-1.5ex}
\begin{equation}
\pointMetric{q}(a,b)=\left(|x_b-x_a|^q + |y_b-y_a|^q\right)^{1/q} = \|a-b\|_q
\label{eq_pointWise_metric}
\end{equation}
\vspace{-3.5ex}

\noindent
By convention, $\pointMetric{q}(a, b)$ is set to zero 
if both $a$ and $b$ exactly lie on the diagonal ($x_a = y_a$ and $x_b = y_b$).
The \jules{$L_q$}-Wasserstein distance, noted 
$\wasserstein{q}$, between $\diagram(f)$ and 
$\diagram(g)$ can then be introduced as:
%
%
\begin{equation}
    \wasserstein{q}\big(\diagram(f), \diagram(g)\big) = 
\min_{\phi
\in \Phi} \left(\sum_{a \in \diagram(f)} 
\pointMetric{q}\big(a,\phi(a)\big)^q\right)^{1/q}
\label{eq_wasserstein}
\end{equation}

\noindent
where $\Phi$ is the set of all possible assignments $\phi$ mapping each 
point
$a \in \diagram(f)$ to 
a point
$b 
\in \diagram(g)$
or to 
its projection onto the diagonal.
$\wasserstein{q}$ can be computed 
via
assignment optimization, for which 
exact \cite{Munkres1957} and approximate \cite{Bertsekas81, Kerber2016}
implementations are publicly available \cite{ttk17}.
}

 \section{Progressive Data Representation}
\label{sec_progressiveData}
This section details our hierarchical scheme for the progressive 
representation of the input \julienRevision{data, which relies on 
a hierarchy of triangulations $\hierarchy$ derived from established 
subdivision schemes~\cite{freudenthal42, bank83, loop87, 
bey95, zhang95}}.
\julienMajorRevision{In particular, our goal is to define a hierarchical scheme 
that will enable efficient update mechanisms between hierarchy levels. This 
will 
avoid, at each new level of the hierarchy, the recomputation
from scratch of the topological data representations presented in 
sections~\ref{sec_progressiveCriticalPoints} and 
\ref{sec_progressivePersistenceDiagram}, and this will instead enable their 
progressive update.}
After a generic description \julienRevision{of the employed triangulation 
hierarchy (\autoref{sec_multiRes})},
we 
\julesReplace{\julienRevision{detail for completeness}}{present for completeness}
an efficient implementation\julienRevision{~\cite{kuhn60, bey98}} for the 
special case of triangulations 
of 
regular grids \julienRevision{(\autoref{sec_gridHierarchy})}, on which we focus 
in this paper (\autoref{sec_limitations} 
discusses generalizations).
Next, we resume our generic 
description \julienRevision{(\autoref{sec_topologicalInvariants})} and 
\julienReplace{show how to leverage the specific structure of the employed 
triangulation hierarchy to accelerate the topological analysis of the data.}
{investigate how the specific structure of the 
employed triangulation hierarchy can be leveraged to accelerate the topological 
analysis of scalar data.} 
For this, we introduce the \julienRevision{novel} notion 
of  \emph{Topologically Invariant Vertices}, \julienRevision{which is} central 
to our \julienRevision{work}.

%

\subsection{Edge-Nested Triangulation Hierarchy}
\label{sec_multiRes}
\begin{figure}
  \center
  \includegraphics*[width=0.9\linewidth]{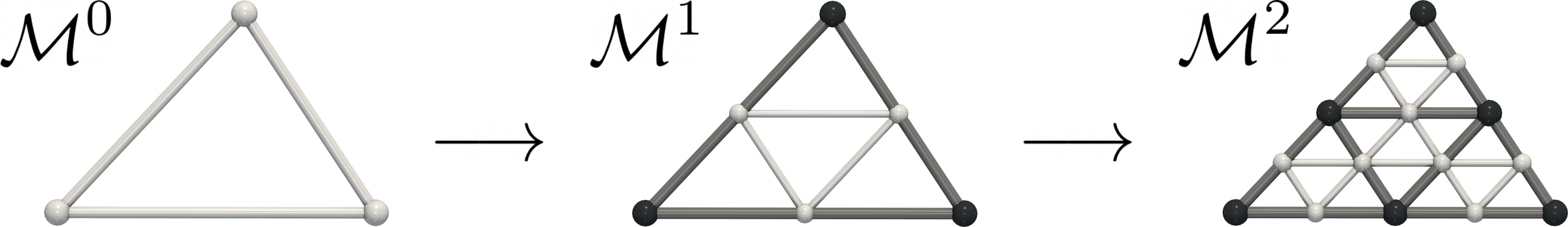}
\mycaption{\julesRevision{Edge-nested triangulation hierarchy 
\julienReplace{for a simple 2D example.}{generated from a 2D triangulation
  constructed using two successive \textit{red} subdivisions of a triangle.}
Old vertices/edges are shown in black/gray. New vertices and
edges are \julienReplace{shown}{showed} in white.} }
\label{fig_edge_nested}
\end{figure}

\julien{Our progressive representation of the input data is based on a 
multiresolution hierarchy of the input PL-manifold $\domain$, 
\julienRevision{which relies on established subdivision schemes 
~\cite{freudenthal42, bank83, loop87, 
bey95, zhang95}.}}
\julesRevision{\julienReplace{Intuitively, our}{The} goal is to 
\julienReplace{define}{provide} a multiresolution hierachy that 
\julienReplace{will enable the efficient update of the topological 
information computed at the previous levels, in order to avoid 
full re-computations (\autoref{sec_progressiveCriticalPoints}).}{allows
topological information
computed on previous levels to remain valid and avoid recomputations 
(\autoref{sec_cpUpdates}).}
In order to construct such a hierarchical scheme, \julienMajorRevision{as 
formalized 
next,} we impose that, as one progresses 
down the 
hierarchy, new vertices are \julienMajorRevision{only} inserted along 
\julienMajorRevision{pre-existing} edges 
(exactly one new vertex per edge, typically \julesReplaceMinor{in}{at} their center), 
and that the additional new edges only connect new vertices 
(\autoref{fig_edge_nested}). 
\julienReplace{This will have the beneficial effect of
preserving, from one hierarchy level to the next, 
the \emph{structure} of the 
local neighborhood around each pre-existing vertex (of its link, as discussed 
in \autoref{sec_topologicalInvariants}), 
which will in turn effectively enable fast updates of the pre-existing local 
topological information (\autoref{sec_progressiveCriticalPoints}).}{This allows 
to preserve the structure of the link of vertices from a level to another.}
We call such a hierarchy \emph{edge-nested} and we formalize 
it in the following, to introduce the notations that will be used in 
the 
rest of the paper.}
\julien{Let $\hierarchy = \{\domain^0, \domain^1, \dots, \domain^h\}$ be a 
hierarchy of PL $d$-manifolds, which respects the following key 
conditions.}

\begin{enumerate}[leftmargin=1.5em]
    \item{\textbf{Old Vertex Condition:} Each vertex of $\domain^i$ (the 
triangulation 
at level $i$)
    also belongs to 
the vertex set, noted $\domain^{i+1}_0$, of $\domain^{i+1}$:
      \begin{eqnarray}
        \domain^i_0 \subset \domain^{i+1}_0
        \label{eq_oldVertices}
      \end{eqnarray}

      The vertices of $\domain^{i+1}$ 
already present in $\domain^{i}$
are called \emph{old vertices} \julienMajorRevision{(black spheres in 
\autoref{fig_edge_nested})}.
    }
    \item{\textbf{New Vertex Condition:} Each vertex of $\domain^{i+1}$
    not present in
     $\domain^{i}$
has to be located 
on an edge $(v_0, v_1)$ of $\domain^i$ (typically \julienReplaceMinor{in}{at} 
its center), 
as
summarized below, where $\domain^{i}_1$ stands for the edge set of 
$\domain^{i}$:
      \begin{eqnarray}
        \forall v \in \domain^{i+1}_0, v \notin \domain^{i}_0 : &
          \exists (v_0, v_1) \in \domain^{i}_1, ~ v \in (v_0, v_1)
      \end{eqnarray}
      
      The vertices of $\domain^{i+1}$ not present in $\domain^{i}$
      are called \emph{new vertices} \julienMajorRevision{(white spheres in 
\autoref{fig_edge_nested})}.
    }
    \item{\textbf{Old Edge Condition:} Each edge $(v_0, v_1)$ of $\domain^{i}$ 
        has to be subdivided at level $i+1$ \julesReplaceMinor{along}{at} exactly one new vertex $v$ of 
$\domain^{i+1}$:
      \begin{eqnarray}
        \begin{aligned}
        \forall (v_0, v_1) \in \domain^{i}_1:  \quad \quad
      & 
         |\{ v \in (v_0, v_1), ~ 
            v \notin \domain^{i}_0, ~
            v \in \domain^{i+1}_0 \}| = 1\\
        & 
        (v_0, v) \in \domain^{i+1}_1, \quad (v, v_1) \in  \domain^{i+1}_1 \\
        & (v_0, v_1) \notin \domain^{i+1}_1
          \end{aligned}
      \end{eqnarray}
      
      The edges of $\domain^{i+1}$ 
      obtained by subdivision of an edge of $\domain^{i}$ are called \emph{old 
      edges}\julesRevision{, they connect old vertices to new vertices} 
\julienMajorRevision{(gray cylinders in \autoref{fig_edge_nested})}.
    }
    \item{\textbf{New Edge Condition:} Each edge of $\domain^{i+1}$ which is 
not an old edge has to connect two new vertices, and it is called a \emph{new 
edge} \julienMajorRevision{(white cylinders in \autoref{fig_edge_nested})}.
    }
%
\end{enumerate}

\julesEditsOut{Overall, 
\julienRevision{this} hierarchical scheme imposes that, as one progresses 
down the 
hierarchy, new vertices are inserted along edges 
(exactly one new vertex per edge, typically in their center), 
and that the additional new edges only connect new vertices 
(\autoref{fig_multires_grid}).}
\julesReplaceMinor{\julienMajorRevision{Figure~\ref{fig_edge_nested}}}{\autoref{fig_edge_nested}} presents a simple example of 
2D edge-nested triangulation hierarchy. Note that the
\julienRevision{Loop subdivision \cite{loop87} 
is compatible with the above formalization, which is more generally termed as 
\emph{red} subdivision in the scientific computing literature, and which has 
been extensively studied for domains of 
two \cite{bank83}, three \cite{bey95, zhang95} and arbitrary dimensions 
\cite{freudenthal42}.}
An input PL manifold $\domain$ admits an edge-nested 
triangulation hierarchy if there exists a hierarchy $\hierarchy$ 
for which $\domain$ is the last element ($\domain = \domain^{h}$).


\begin{figure}[b]
  \centering
  \includegraphics[width=\linewidth]{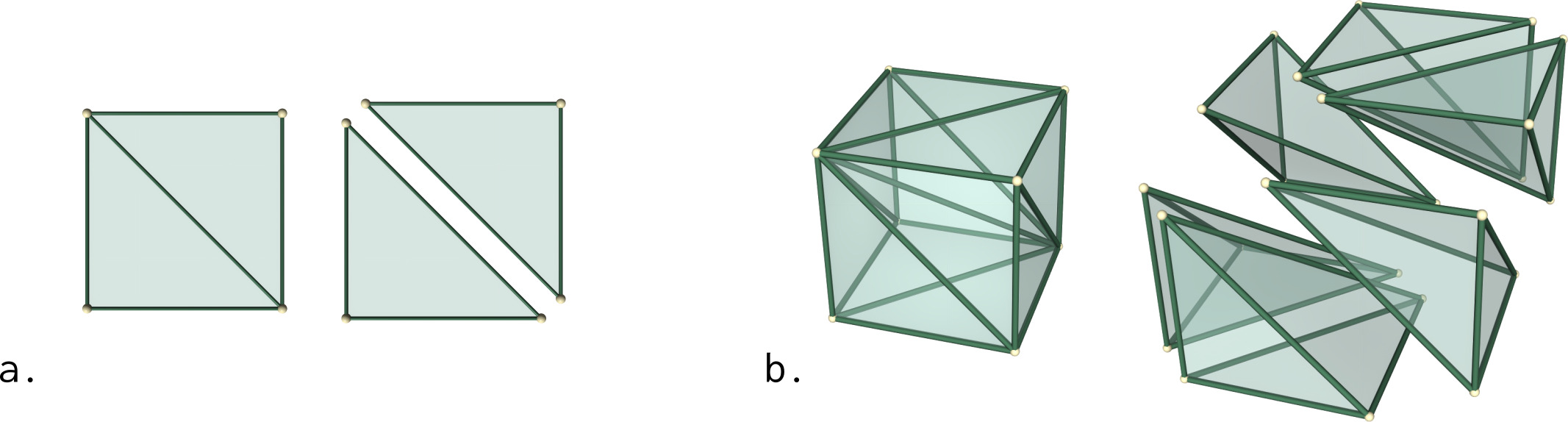}
  \mycaption{Translation invariant local triangulation pattern for the cells of 
a 
2D and 3D regular grid.
    In 2D, quadrilaterals are subdivided into two triangles (a), always 
along the same diagonal. In 3D, the generalization of this pattern 
subdivides each hexahedron into six tetrahedra (b).
}
  \label{fig_triangulatingGrid}
\end{figure}

\subsection{Edge-Nested Triangulations of Regular Grids}
\label{sec_gridHierarchy}

\julienRevision{While the construction of an edge-nested triangulation 
hierarchy given an arbitrary 
input manifold $\domain$ is an open question which we leave for future work 
(see \autoref{sec_limitations}), it can be shown that such a hierarchy 
exists for regular grids, and that it can be 
implemented 
very
efficiently, as discussed by Bey~\cite{bey98}. We describe this implementation 
in the following for the sake of completeness\julienReplace{, by detailing how 
to efficiently retrieve an 
arbitrarily coarse version of the fine triangulation $\domain^{h}$ from an 
input 
regular grid $\Grid^{0}$.}{.}}

Let $\Grid^{0}$ be a $d$-dimensional regular grid, with $d$ \julesReplace{equals}{equal to} $2$ or $3$
in our applications, of dimensions $\xDim^{0}$, $\yDim^{0}$, $\zDim^{0}$ 
$\big(\text{\julesRevision{\textit{i.e.} of number of vertices }}|\Grid^{0}_0| = (\xDim^{0} + 1)\times (\yDim^{0}+1) \times 
(\zDim^{0}+1)$, in 2D: $\zDim^{0} = 0\big)$.
We will 
first 
assume that $\xDim^{0}$, $\yDim^{0}$ and $\zDim^{0}$ are all powers of $2$. 
Let $\triangulationOp_0$ be the
\emph{triangulation operator},
which transforms 
$\Grid^{0}$ into a valid triangulation $\domain^{h}$, i.e. $\domain^{h} = 
\triangulationOp_0(\Grid^{0})$, 
by preserving vertex sets, i.e. $\domain^{h}_0 = \Grid^{0}_0$, and 
by inserting exactly one edge for each $i$-dimensional cell of  $\Grid^{0}$ 
($1 < i \leq d$),
%
according to a unique pattern, which is \emph{invariant by 
translation} along the cells of the grid\julienRevision{, known as Kuhn's 
triangulation \cite{kuhn60}}.
In $2$D, each 
quadrilateral is subdivided into two triangles by inserting one edge 
always along the \emph{same diagonal}. In $3$D, each 
hexahedron is subdivided into six tetrahedra by 
always inserting the \emph{same diagonal} edges
\julien{(\autoref{fig_triangulatingGrid})}.

\begin{figure}
\begin{center}
\adjustbox{width=\linewidth,center}{
    \begin{tikzcd}
        \centering
          \domain^{0} \arrow[r]  
            & \domain^{1} \arrow[r] 
            & \dots \arrow[r]  
            & \domain^{h - 1} \arrow[r]
            & \domain^{h}\\
         \Grid^{h} \arrow[u, "\triangulationOp_{h}"]
            & \Grid^{h-1} \arrow[l, , "\decimationOp_{h}"] \arrow[u, 
"\triangulationOp_{h - 1}"]
            & \dots \arrow[l, "\decimationOp_{h - 1}"]  \arrow[u]
            & \Grid^{1} \arrow[l, "\decimationOp_2"] \arrow[u, 
"\triangulationOp_{1}"]
            & \Grid^{0} \arrow[l, "\decimationOp_1"] \arrow[u, 
"\triangulationOp_{0}"]
    \end{tikzcd}
    }
    \mycaption{Commutative diagram for the generation of an edge-nested
triangulation hierarchy $\hierarchy = \{\domain^0, \domain^1, \dots, 
\domain^h\}$ from a regular grid $\Grid^0$. The hierarchy can be obtained by a 
sequence of decimation operators $\decimationOp_i$, accompanied with 
triangulation 
operators $\triangulationOp_i$.}
    \label{fig_commutativeDiagram}
\end{center}
\end{figure}

Let $\decimationOp_1$ be the \emph{decimation operator}, which
transforms the regular grid $\Grid^{0}$ into a regular grid $\Grid^{1}$, i.e.  
$\Grid^{1} = \decimationOp_1(\Grid^{0})$, 
by selecting one vertex every 
two vertices in each dimension. 
Let $(i, j, k)$ be the grid coordinates of a vertex $v \in \Grid^{0}$. 
Then the grid $\Grid^{1}$ is obtained by only selecting the vertices with even 
grid coordinates $(i, j, k)$ in $\Grid^{0}$. 
In $2$D, each 
quadrilateral of $\Grid^{1}$  corresponds in the general case to four
quadrilaterals of $\Grid^{0}$ 
and in $3$D, each hexahedron of $\Grid^{1}$ corresponds to eight hexadra of 
$\Grid^{0}$. Note that the decimation operator $\decimationOp_1$ induces a 
reciprocal 
\emph{subdivision}
operator, which, given  $\Grid^{1}$, yields $\Grid^{0}$ by 
inserting a new 
vertex in the center of each $i$-dimensional cell of $\Grid^{1}$ ($0 < i \leq 
d$).

\begin{figure}
  \center
  \includegraphics*[width=0.9\linewidth]{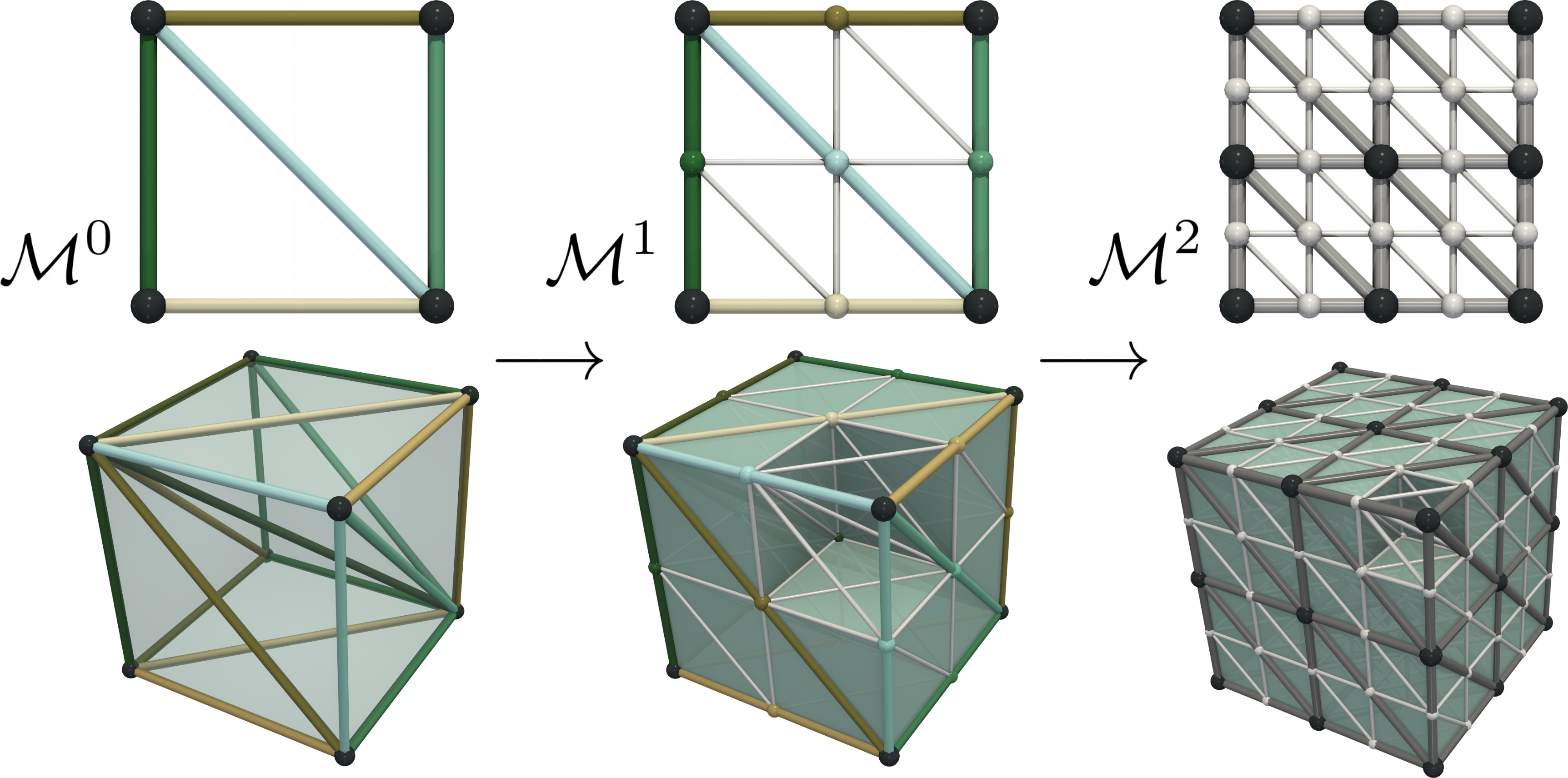}
\mycaption{Edge-nested triangulation hierarchy generated from a regular grid.
Old vertices/edges are shown in black/gray in $\domain^2$.
There is a one-to-one mapping  (colors from $\domain^0$ to $\domain^1$)
between the edges of $\domain^0$
and the new 
vertices of $\Grid^{h-1}$, inserted 
in
each $i$-dimensional cell 
of $\Grid^{h}$ ($0 < i \leq d$).}
\label{fig_multires_grid}
\end{figure}

We now introduce by recurrence a sequence of decimation operators 
$\decimationOp_i$ 
(\autoref{fig_commutativeDiagram}), 
which decimate each grid $\Grid^{i-1}$ into a grid $\Grid^{i}$ by sub-sampling 
its vertices with even grid coordinates as described above. It follows that for 
a given level of decimation $i$, the dimensions of $\Grid^{i}$ are given by 
$\xDim^{i} = \xDim^{0}/2^{i}$,
$\yDim^{i} = \yDim^{0}/2^{i}$, and
$\zDim^{i} = \zDim^{0}/2^{i}$.
Let us now consider the sequence of triangulation operators 
$\triangulationOp_i$, which 
triangulate each grid $\Grid^{i}$ into a triangulation $\domain^{h - i}$, i.e. 
$\domain^{h - i} = \triangulationOp_i(\Grid^{i})$, as illustrated by the 
commutative 
diagram of \autoref{fig_commutativeDiagram}. 
Then, it can be verified (\autoref{fig_multires_grid}) that each 
condition of \autoref{sec_multiRes} is indeed satisfied by the sequence 
$\hierarchy = \{\domain^0, \domain^1, \dots, \domain^h\}$ and that $\hierarchy$ 
is  a valid edge-nested triangulation hierarchy.
\julienRevision{In particular, as described by Bey~\cite{bey98}, any 
triangulation $\domain^{i}$ can be equivalently obtained either: \emph{(i)} by 
applying the red subdivision scheme \cite{bank83, bey95, zhang95} $i$ times on 
$\domain^0$ or \emph{(ii)} by considering the Kuhn 
triangulation~\cite{kuhn60} 
of $\Grid^{h-i}$ (itself obtained by $i$ regular subdivisions of $\Grid^{h}$). 
In other words, any triangulation $\domain^{i}$ in the commutative diagram 
of \autoref{fig_commutativeDiagram} can be obtained by starting either 
\emph{(i)} from $\domain^0$ or \emph{(ii)} from  $\Grid^h$. In our work, we 
exploit this equivalence property, but in \emph{reverse}: we 
use it to efficiently retrieve an arbitrarily coarse version of 
the fine triangulation $\domain^h$ of the input grid $\Grid^0$.}

\julienRevision{In particular, the}
edge-nested triangulation hierarchy 
\julienRevision{$\hierarchy$}
can be implemented very 
efficiently, by encoding the entire hierarchy implicitly, and by only 
maintaining the grid  $\Grid^{0}$ in memory. At a given hierarchy level $i$, 
adjacency relations in $\domain^i$ between two vertices $v_0$ and $v_1$ can be 
inferred based on their grid coordinates at level $i$, $(i_0, j_0, k_0)$ and 
$(i_1, j_1, k_1)$, and given the triangulation pattern shown in 
\autoref{fig_triangulatingGrid}. Then, the data values associated to the 
vertices $v_0$ and $v_1$ can be retrieved by mapping these vertices back to 
their original locations in $\Grid^{0}$, given by the grid coordinates 
$(i_0\times2^{h-i}, j_0\times2^{h-i}, k_0\times 2^{h-i})$ 
and $(i_1\times2^{h-i}, j_1\times2^{h-i}, k_1\times 2^{h-i})$. 
This approach is easily extended to support regular grids whose 
dimensions, 
$\xDim^{0}$, $\yDim^{0}$ or $\zDim^{0}$
are not necessarily powers of 2. 
In particular, when considering the decimation operator $\decimationOp_i$,
in case some of the dimensions  
$\xDim^{i-1}$, $\yDim^{i-1}$ or $\zDim^{i-1}$ 
are not even, 
$\decimationOp_i$ 
systematically 
adds the last vertex of $\Grid^{i-1}$ for each odd dimension.
\julien{In our progressive algorithms (Sec.~\ref{sec_progressiveCriticalPoints} 
and 
\ref{sec_progressivePersistenceDiagram}), these few extra vertices will require
full recomputations.}
Below,
we resume our generic description for arbitrary
edge-nested triangulation hierarchies, not necessarily obtained from regular 
grids (\autoref{sec_limitations} discusses generalizations).

\begin{figure*}
  \centering
  \includegraphics[width=\linewidth]{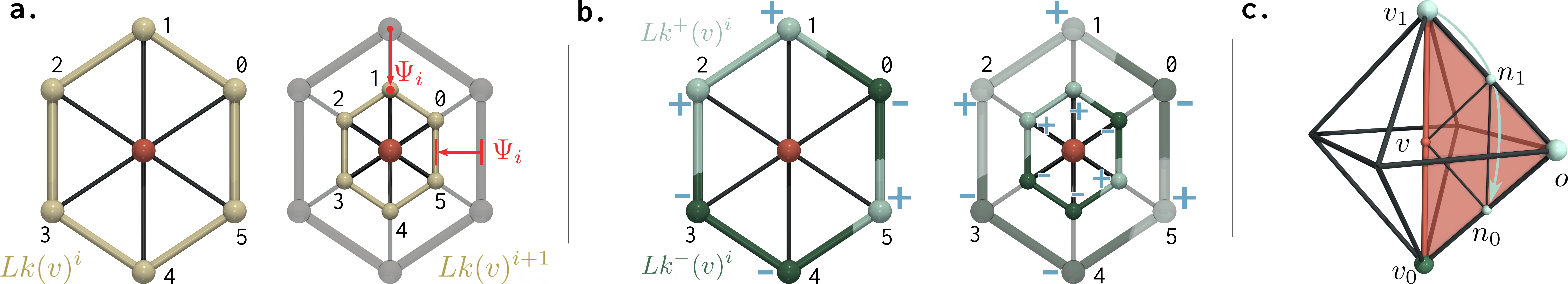}
  \mycaption{\julienMajorRevision{Important properties of edge-nested 
triangulations, enabling fast updates of local topological 
information. (a) Left: From one hierarchy level ($i$) to the next ($i+1$), 
edge-nested triangulations preserve the local structure of the link 
$\Link(v)^i$ of an old vertex $v$ (red sphere). In particular, there exists a 
one-to-one mapping $\Psi_i$ between the vertices and the edges (red arrows) of 
$\Link(v)^i$ and $\Link(v)^{i+1}$. (b) Center: this link invariance enables 
the fast identification of old vertices which do not change their
\julesReplaceMinor{critical type}{criticality}: these are old vertices (red
sphere) for which the \emph{polarity} (blue 
signs) remains unchanged from one hierarchy level ($i$) to the next ($i+1$) and 
for which, therefore, connected components of lower and upper links (green and 
blue components, respectively)  do not change (thus, requiring no 
update). Such vertices are called 
\emph{topologically invariant old vertices}. (c) Right: A new  
vertex $v$ which is \emph{monotonic} (i.e. $f(v_0) < f(v) < f(v_1)$, with $v_0$ 
and $v_1$ being respectively the lowest and highest vertex of the edge $(v_0, 
v_1)$ where $v$ is inserted) is guaranteed to be regular 
if all its adjacent new neighbors (in the figure, $n_0$ and $n_1$) are also 
\emph{monotonic} (see \autoref{sec_topologicalInvariants} for further 
discussion).}}
  \label{fig_topo_invariants}
\end{figure*}

\subsection{Topologically Invariant \julien{Vertices}}
\label{sec_topologicalInvariants}



The input edge-nested triangulation hierarchy $\hierarchy$ 
yields a hierarchy of PL scalar fields $\{f^{0}, f^{1}, 
\dots, f^{h}\}$, 
such that each old vertex $v$ maintains by 
construction its scalar value:
$f^{i}(v) = f^{j}(v) 
= f(v), ~ \forall j ~ / ~ i \leq j \leq h$.
In the following, we show how the
specific structure of 
edge-nested triangulation 
hierarchies 
\julesRevision{described in \autoref{sec_multiRes}}
can be leveraged 
to 
efficiently update topological information while progressing down the hierarchy.
First we 
show that edge-nested triangulations 
preserve the topology of 
the link
of vertices when 
progressing from one hierarchy level to the next. 
This enables the quick identification, discussed next, of 
  vertices which do not change their \julesReplaceMinor{critical type}{criticality} when progressing down the 
  hierarchy\julesReplaceMinor{, and which we call \emph{topologically invariant old vertices} and 
  for which no update will be needed during subsequent analyses }{. We call these
  vertices \emph{topologically invariant old vertices}, as they will need
  no update during subsequent analyses}
\julesReplaceMinor{(sections~\ref{sec_progressiveCriticalPoints} and
\ref{sec_progressivePersistenceDiagram})}{(\autoref{sec_progressiveCriticalPoints} and \autoref{sec_progressivePersistenceDiagram})}.
\julienMajorRevision{Last, we show how to efficiently identify new vertices 
that are guaranteed by construction to be regular points of $f^i$, which we 
call \emph{topologically invariant new vertices} and for which no computation 
will be required in subsequent analyses.}

\noindent
\textbf{1) Link Topological Invariance:}
A first  insight 
is that 
the link $\Link(v)$ of a vertex $v$ is topologically 
invariant throughout the hierarchy.
\julienMajorRevision{This property is important because it will enable the 
fast identification of vertices which do not change their 
\julienReplaceMinor{critical}{criticality}
(next 
paragraph).}
Let $\Link(v)^{i}$ be the link of 
$v$ at level $i$, then there exists a 
one-to-one mapping 
$\Psi_i$ (\julesReplace{right inset}{\autoref{fig_topo_invariants}(a)})
between the simplices of $\Link(v)^{i}$ and 
$\Link(v)^{i+1}$ --
such that $\Link(v)^{i+1} = \Psi_i\big(\Link(v)^{i}\big)$ --
which preserves the simplicial structure of $\Link(v)^{i}$
(which preserves adjacencies). 
Indeed,
\emph{(i)} new vertices are only inserted on old edges (this maps the
$k^{th}$ neighbor of $v$ at level $i$ to its $k^{th}$ new neighbor at level
$i+1$, \julienMajorRevision{top red arrow in \autoref{fig_topo_invariants}(a)}) 
and \emph{(ii)} new edges are only inserted between new
vertices
(this maps the $k^{th}$ edge of $\Link(v)^{i}$ to the
$k^{th}$ new edge of $\Link(v)^{i+1}$, 
\julesReplace{inset}{\julienMajorRevision{right red 
arrow in }\autoref{fig_topo_invariants}(a)}). This
mapping $\Psi_i$ can be 
viewed as a 
combinatorially
invariant \emph{zoom} in the neighborhood of $v$ as one 
progresses down the hierarchy. 


\noindent
\textbf{2) Topologically Invariant Old Vertices:}
A second insight deals with the evolution of the data values on the link 
of an \emph{old} vertex, as one 
progresses down the hierarchy and zooms with the above mapping $\Psi_i$.
We define the \emph{polarity} of $\Link(v)^{i}$, noted 
$\delta : \Link(v)^{i} \rightarrow \{-1, 1\}$ as the field which assigns 
to 
each neighbor $n$ of $v$ at level $i$ the sign of its function difference with 
$v$: 
$\delta(n) = sgn\big(f(n) - f(v)\big)$. 
The polarity is positive 
in the upper link,  negative 
in
the lower link 
\julesReplace{(blue signs, above inset)}{(\autoref{fig_topo_invariants}(b), 
blue signs)}.
Let 
$(v_0, v_1)$ be an 
edge at level $i$, which gets subdivided at level 
$i+1$ 
along a new vertex $v_n$. \julesReplaceMinor{W}{Assuming that $f(v_0)<f(v_1)$}, we say that $v_n$ is \emph{monotonic} if 
\julesReplaceMinor{ $f(v_0) < f(v_n) < f(v_1)$ or $f(v_0) > f(v_n) > f(v_1)$}{
$f(v_n) \in \big(f(v_0),f(v_1)\big)$}.
Otherwise, 
$v_n$ is  \emph{non-monotonic}. 
In that case, if $v_n$'s polarity in $\Link(v_0)^{i+1}$ is the opposite of 
$v_1$'s polarity in $\Link(v_0)^{i}$, 
we say that $v_0$ is \emph{impacted} by its 
neighbor $v_n$.
Now, if an old vertex $v$ is not \emph{impacted} by any of its non-monotonic 
neighbors,
its link polarity is maintained \julienReplace{(i.e. the blue signs in 
\autoref{fig_topo_invariants}(b) remain unchanged when going from the 
hierarchy level $i$ to $i+1$). This implies that $v$ is therefore guaranteed 
to maintain its \emph{criticality}:}{
\julesReplace{(above inset)}{(\autoref{fig_topo_invariants}(b)} and
$v$ is guaranteed to maintain its \emph{criticality}:}
it maintains its critical index (i.e., $\Index(v)^{i+1} = \Index(v)^{i}$) 
or it remains regular.
\julienRevision{Indeed, each 
  neighbor $n$ which does not impact $v$ maintains \julesReplaceMinor{it}{its}
classification as being upper or lower.
Then,
since there is a one-to-one 
mapping $\Psi_{i}$ (see \julesReplace{the inset figure in the above 
paragraph}{\autoref{fig_topo_invariants}(a)}) between 
$\Link(v)^{i}$
and
$\Link(v)^{i+1}$ 
which preserves their simplicial structure, 
it follows that the   complexes 
$\lowerlink(v)^{i+1}$ and
$\upperlink(v)^{i+1}$ are 
respectively identical to
$\lowerlink(v)^{i}$ and
$\upperlink(v)^{i}$.  Thus, the number of connected components of lower and
upper links are maintained, preserving the criticality of $v$.}
Old vertices which are not impacted by their 
non-monotonic
neighbors
are called \emph{topologically 
invariant old vertices}.

\noindent
\textbf{3) Topologically Invariant New Vertices:}
A third insight deals with the link of \emph{new} vertices.
Given a new monotonic vertex $v$ \julienMajorRevision{(small red 
sphere
in \autoref{fig_topo_invariants}(c))} subdividing an edge 
\julienMajorRevision{($v_0, v_1$)}
at level $i$ \julienMajorRevision{(red cylinder in  
\autoref{fig_topo_invariants}(c))}, 
if its new neighbors are all monotonic as well, $v$ is 
then called an \emph{interpolating vertex} and it can be shown that $v$ 
must
be a regular vertex.
\julienRevision{First, since $v$ is monotonic, it cannot 
be an 
extremum, since by definition it is connected to one lower ($v_0$) and one 
upper ($v_1$) old vertex 
\julienMajorRevision{(large green and blue spheres 
in \autoref{fig_topo_invariants}(c))}. Note that $v_0$ and $v_1$ are the only 
old vertices 
adjacent to $v$.
Second, to show that $v$ is regular, we argue
that
$\upperlink(v)^{i}$ is necessarily connected (and so is 
$\lowerlink(v)^{i}$, symmetrically). Let $(v_0 ,v_1, o)$ be a triangle at level 
$i-1$  (\julesReplace{above 
inset}{\julienMajorRevision{red triangle in 
}\autoref{fig_topo_invariants}(c)}). At level $i$, the edges $(v_0, o)$ 
and $(v_1, o)$ are subdivided 
along the new vertices $n_0$ and 
$n_1$ and 
the \emph{new} edges $(v, n_0)$, $(v, n_1)$, and $(n_0, n_1)$ are inserted to 
connect the new vertices.
Let us assume 
that $f(n_0) > f(v)$. $n_0$ is then an \emph{upper} neighbor of $v$ ($n_0 
\in \upperlink(v)^{i}$). Since $n_0$ is monotonic, this means that the 
\emph{outer} old vertex $o$  \julesReplaceMinor{}{(which is not in 
$\Link(v)^i$)}
must also be
upper: $f(o) > f(n_0) > f(v)$. 
Since $n_1$ is monotonic as well, it follows that $n_1$ is upper too.
Thus, there exists a path 
$
\{v_1, n_1, n_0\}
\in \Link(v)^{i}$ \julienMajorRevision{(blue arrow in  
\autoref{fig_topo_invariants}(c))}, which connects 
$v_1$ to $n_0$ 
and which is only composed of \emph{upper} vertices. Thus 
$n_0$ and 
$v_1$ belong to the same connected component of $\upperlink(v)^{i}$. The same 
reasoning holds for all the new \emph{upper} neighbors of $v$.  It follows that 
$\upperlink(v)^{i}$ and 
$\lowerlink(v)^{i}$ are both made of a single connected component, 
containing exactly one old vertex each, $v_1$ and $v_0$ respectively.
Thus, $v$ is regular.
Note that this reasoning readily applies to 2D and 3D. 
}
Since interpolating vertices, such as $v$, imply no topological event in the 
sub-level sets, 
we call them \emph{topologically invariant new
vertices}.

%

\begin{figure}[t]
  \includegraphics[width=\linewidth]{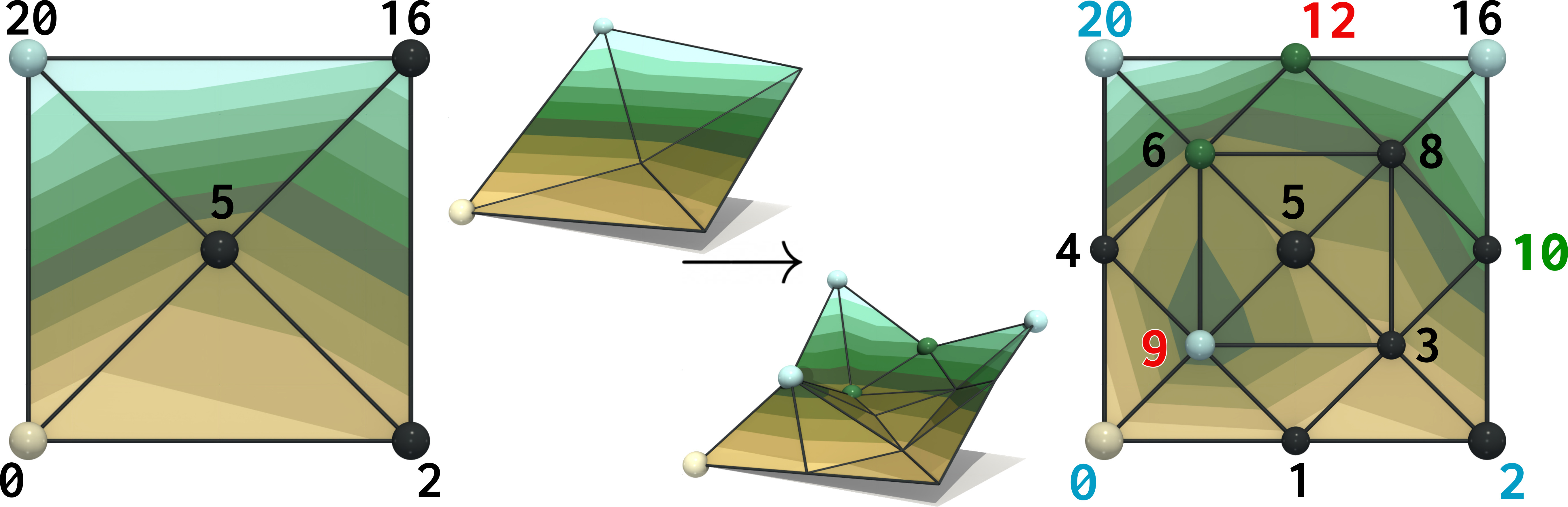}
  \mycaption{\emph{Topologically Invariant} (TI) vertices (numbers denote $f$ 
values). 
When progressing down the hierarchy, two \emph{non-monotonic} vertices appear 
(red labels). This yields new critical points (cyan: maxima, green: saddles, 
brown: minimum).
\textit{Old} TIs (blue labels), whose link polarity is unchanged, 
maintain their
criticality. 
\textit{New} TIs are regular (green label).
\julienMajorRevision{For \emph{topologically invariant} 
vertices (blue and green labels), no computation is required. As illustrated 
in \autoref{tab_stats}, TI vertices represent the majority of the data 
in real-life datasets.}}
\label{fig_monotony_change}
\end{figure}

The three key insights of edge-nested triangulations discussed above 
(summarized in \autoref{fig_monotony_change}) form the 
cornerstone of our progressive approach to topological analysis. 
As detailed next,
checking if 
vertices are 
topologically invariant \julesReplaceMinor{reveals}{turns out} to be  less computationally expensive in 
practice than 
computing their criticality from scratch.
Moreover, the set of topologically invariant vertices 
tend\julesReplaceMinor{}{s} to represent \julesReplaceMinor{in practice }{}the majority of the hierarchy (see 
\autoref{sec_results}).
This allows for the design of efficient 
progressive algorithms, presented in the next sections.

\section{Progressive Critical Points}
\label{sec_progressiveCriticalPoints}
Our progressive algorithm for critical point extraction
\julesReplace{initializes}{starts} at the first level of the hierarchy, 
$\domain^0$, and 
progresses  
level by level 
down the hierarchy $\hierarchy$
until reaching 
its final level,
$\domain^h$, or 
until \julesReplaceMinor{user interruption}{interrupted by a user}.
\julienMajorRevision{At each level $i$, our approach delivers the 
entire list of critical points of the data for the current resolution ($f^i : 
\domain^i \rightarrow \range$). For this, our strategy consists in avoiding 
recomputation as much as possible and instead efficiently and minimally 
update the information computed at the previous level $(i - 1)$.}


\subsection{Initialization and Updates}
\label{sec_cpUpdates}
%
%
%

This section focuses on
the vertices of $\hierarchy$ which are not \emph{topologically 
invariant}.
The 
case of topologically 
invariant vertices is discussed in \autoref{sec_cpRegularity}.
\julienReplaceMinor{}{In short, our approach computes the 
criticality of each vertex with the traditional method \cite{banchoff70} at 
the first hierarchy level. However, for the following levels, instead of 
re-starting this computation from scratch, our algorithm maintains the 
criticality information computed at the previous levels and only 
minimally updates this information, where needed, by using  dynamic trees 
\cite{sleator83}, a specialized data structure for dynamic connectivity 
tracking.}

At the first hierarchy level, 
$\domain^0$ 
only 
contains new vertices for which the criticality needs to be initialized. As of 
the second level, old and new vertices start to co-exist 
in  $\domain^1$
and fast update mechanisms can be considered to efficiently update the 
criticality of the old vertices.
For this, we leverage the 
topological invariance of the link of each old vertex
throughout the hierarchy (\autoref{sec_topologicalInvariants}). 
This allows to 
store relevant topological information on the link and to quickly update them 
when progressing down the hierarchy. In particular, we initialize for each 
\emph{new} vertex 
$v$ 
at \julesReplaceMinor{the }{}level $i$ the following information:
\begin{itemize}[leftmargin=.85em]
  \item{\emph{Link 1-skeleton:} We store the list of 
\emph{local} edges (and their adjacencies) of
$\Link(v)^{i}$\julesReplaceMinor{.}{, encoded with pairs of local indices for the
neighbors of $v$.}
This remains invariant through $\hierarchy$
(\autoref{sec_topologicalInvariants}).}
  \item{\emph{Link polarity:} We store for each vertex of $\Link(v)^{i}$ 
its \emph{polarity} (\autoref{sec_topologicalInvariants}), i.e. its 
classification as being upper or lower than $v$. This is encoded with one 
bit per vertex of $\Link(v)^{i}$.}
  \item{\emph{Link dynamic tree:} An efficient data structure \cite{sleator83} 
for maintaining connected components in dynamic graphs, discussed 
below.}
\end{itemize}

For each new vertex $v$ which is not topologically invariant, 
\julienReplaceMinor{these}{the following}
data 
structures are initialized\julesReplaceMinor{ and t}{: a list of pairs of local neighbor indices denoting the local edges of $\Link(v)^i$ (up to 24 pairs in a 3D grid), 
a list of bits denoting the polarity of each neighbor (up to 14 neighbors in a 3D grid), and the dynamic tree, detailed below. T}he criticality of $v$ is computed with the 
traditional approach (\autoref{sec_criticalPoints}), by enumerating the 
connected components of $\upperlink(v)^{i}$ and $\lowerlink(v)^{i}$. 
This is usually achieved with
breadth-first search traversals or with a Union-Find (UF) data structure 
\cite{cormen}. However, in our setting, we would like to update 
these connected components as the algorithm progresses down the hierarchy. In 
particular, if a \emph{local} edge $e$  belongs to the upper link of $v$ at 
\julesReplaceMinor{the }{}level $i$, but not anymore at \julesReplaceMinor{the }{}level $i+1$, the connected components of 
$\upperlink(v)^{i+1}$ need to be updated accordingly, preferably without 
recomputing them completely. For this, we use dynamic trees \cite{sleator83}, 
which, like the UF data structure, maintain connected components in a graph 
upon edge insertion, but unlike the UF, also maintain them upon edge removal. 
In particular, all the vertices of $\Link(v)^{i}$ are initially inserted in the 
dynamic tree associated to $v$. Next, 
we insert each local edge of $\Link(v)^{i}$
in the dynamic tree,
if both its \julesReplaceMinor{extremities}{ends} \julesReplace{admit}{have} 
the same polarity. \julesReplaceMinor{Then, t}{T}he criticality of 
$v$ is \julesReplaceMinor{}{then }deduced by enumerating the connected components with positive and 
negative polarity, thanks to the dynamic 
tree.

\begin{figure}
  \centering
  \includegraphics[width=\linewidth]{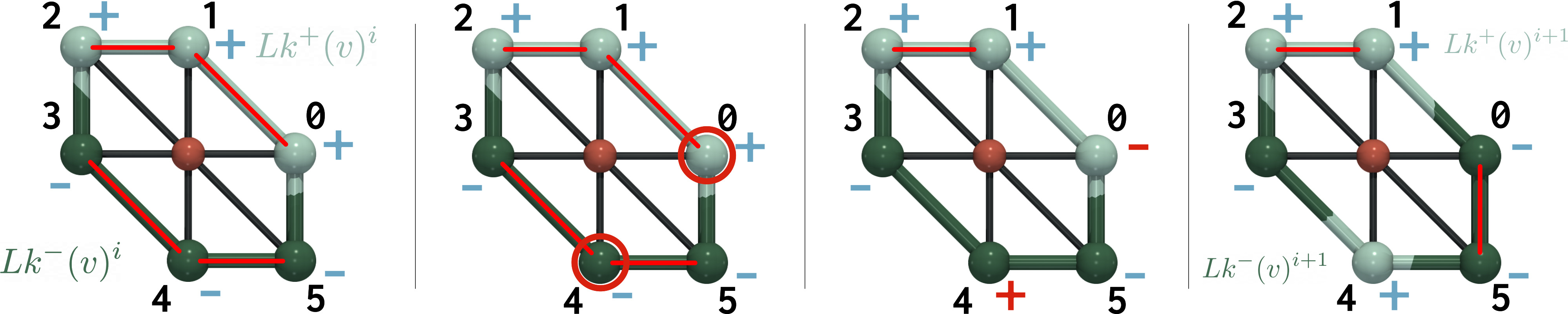}
  \mycaption{Updating the criticality of a non-topologically invariant old 
vertex. From
    left to right: initial state, identification of \textit{non-monotonic} 
vertices (red circles), 
  update of the 
  \textit{link polarity} (red $+$/$-$ signs), and update of the connected 
components
of $\upperlink(v)$ and $\lowerlink(v)$. 
At each step, edges present in the dynamic tree \cite{sleator83} are 
highlighted in red.
Only the edges impacted by polarity flips need to be updated in the 
dynamic tree: edges $(0, 1)$, $(3, 4)$ and $(4, 5)$ are 
removed, and the edge
$(0, 5)$ is added.
}
  \label{fig_dynamic_link_update}
\end{figure}

For each old vertex $v$ which is not topologically invariant 
(\autoref{fig_dynamic_link_update}), its link polarity 
is quickly updated based on the non-monotonic new vertices 
 of $\Link(v)^{i}$.
Each local edge $e$ of $\Link(v)^{i}$ which is 
\emph{impacted} by a polarity flip of 
its vertices (\autoref{sec_topologicalInvariants}) 
is removed from the 
dynamic tree associated to $v$ if it was present in it
(to account for the corresponding disconnection of lower/upper link component),
and 
added to 
it otherwise,
if both its \julesReplaceMinor{extremities}{ends} \julesReplace{admit}{have} the same polarity 
(if they belong to the same  lower/upper link component).
Then, the criticality of $v$ is quickly updated with the 
fast enumeration of the connected components of positive and negative 
polarity provided by the dynamic tree.
\julesReplace{}{\julienReplace{Note that such}{Such} an efficicent update of 
the criticality of $v$ would not
be feasible with a simple UF data structure, as the connected components of the link of $v$
would need to be recomputed from scratch upon edge removal.}
\subsection{Computation Shortcuts}
\label{sec_cpRegularity}


When moving 
from the hierarchy level $i$ to $i+1$, topologically invariant old vertices 
are guaranteed to maintain their criticality 
(\autoref{sec_topologicalInvariants}). 
For these, the dynamic trees 
(\autoref{sec_cpUpdates}) do not need to be updated.
Moreover, when moving from the hierarchy level $i$ to $i+1$, each topologically 
invariant new vertex $v$ 
is guaranteed 
to be regular. 
For these, the dynamic trees (\autoref{sec_cpUpdates}) 
are not even initialized (they will only be used when $v$ becomes no longer 
topologically invariant).
Overall, our procedure to update vertex criticality
can be summarized as follows:

  \noindent
  \textbf{1) Mononotic vertices:} in this step, we loop over 
all new vertices to check whether or not they are monotonic.
  
  \noindent
  \textbf{2) Link polarity:} in this step, we loop over all 
vertices to initialize/update their link polarity. For old vertices, updates are
only needed for their non-mononotic neighbors.
If an old vertex $v$ is topologically invariant, 
no more 
computation is required for it at this hierarchy level.
  
  \noindent
  \textbf{3) Old vertices:} each old vertex $v$ which is not 
topologically invariant efficiently updates 
its criticality in $f^{i}$ as described in \autoref{sec_cpUpdates}.
  
  \noindent
  \textbf{4) New vertices:} if a new vertex $v$ is 
  topologically invariant, it
is classified as regular  and no more 
computation is required for it at this hierarchy level. Otherwise, 
its criticality is updated (\autoref{sec_cpUpdates}).

\subsection{Parallelism}
\label{sec_cpParallel}
Critical point computation is an operation which is local to the link of each 
vertex. 
%
Thus, each of the four steps introduced above 
can be 
trivially parallelized over the vertices of 
$\domain^i$ with shared-memory parallelism. This implies no 
synchronization, at the exception 
of the sequential transition between two consecutive steps.

%

\subsection{Extremum Lifetime}
\label{sec_cpLifeTime}
As our algorithm progresses down $\hierarchy$, 
the 
population of critical points evolves. 
In practice,
this means that some features of interest may 
be captured by the 
progressive algorithm earlier than others, 
denoting
their importance 
in the data.
To evaluate this, we consider for each extremum $e$ the notion of 
\emph{Lifetime}, \julesReplaceMinor{noted}{defined as} $l(e) = l_d(e) - l_a(e)$, 
where $l_a(e)$ and $l_d(e)$ stand for the levels where $e$ appeared 
and disappeared respectively.
The evaluation of this 
measure 
requires 
a 
correspondence between the extrema computed at the 
levels $i$ and $i+1$,
which is 
in general a challenging assignment optimization problem 
\cite{Ji2006a,
KleinE07, 
bremer_tvcg11,
ReininghausKWH12,
SaikiaW17,
soler_ldav18}.
%
For simplicity, we focus here on a simple yet time-efficient heuristic for 
estimating these correspondences, which can be enabled optionally. 

Given a vertex $v$, identified as maximum at hierarchy level $i-1$,
our 
heuristic consists \julesReplaceMinor{in}{of} computing, for each neighbor $n$ of $v$, an 
integral line $\forwardIntegralLine(n)^{i}$.
Each of these lines terminates on local 
maxima of $f^{i}$, which we add to the set of \emph{candidates} for $v$. 
At the end this step, we 
establish 
the correspondence between $v$ and its 
highest candidate in terms of $f^{i}$ values, noted $m^{*}$, and we say that 
$v$ 
\emph{maps} to $m^{*}$ from $i-1$ to $i$. 
To focus the integration on a reasonable neighborhood,
we restrict the number of edges on each integral 
line to a user 
parameter $L_{max}$, 
set to \jules{$10$}
in our experiments. 
If the set of \emph{candidates} of $v$ is empty, 
the maximum present in $v$ at level $i-1$ is considered to disappear at level 
$i$ $\big(l_d(v) = i\big)$. 
It is possible, given a maximum $m$ at 
level $i$, that no maximum from the level $i-1$ maps to it. In this case, $m$ 
is said to appear at the level $i$ $\big(l_a(m) = i\big)$. 
Finally, 
if multiple 
maxima at level $i-1$ map to the same 
maximum at level $i$, they are all 
considered to disappear at the level $i$, at the exception of the \emph{oldest} 
maximum (minimizing $l_a$), 
as 
suggested by the 
\emph{Elder} rule in the case of 
persistence \cite{edelsbrunner09}. 
This optional procedure is run at each hierarchy level and 
enables the progressive estimation of the lifetime of the maxima. 
Note that the lifetime of minima is estimated with the symmetric procedure.

\begin{figure}
        \includegraphics[width=\linewidth]{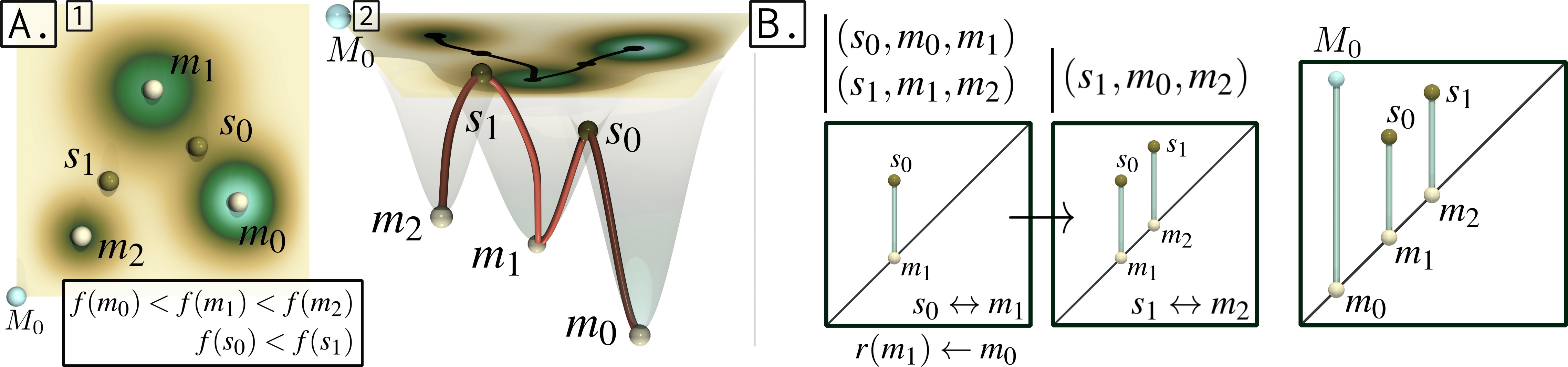}
 \mycaption{
 Computing the minimum-saddle persistence diagram from critical points. 
 Downwards monotonic paths are initiated at saddles to extract a list of 
critical point triplets (part A, left), which forms a reduced topological 
representation of the data. This reduced reprensentation is efficiently 
processed to produce the persistence diagram (part B, right).}
  \label{fig_pd_nonProgressiveStrategy}
\end{figure}

\begin{figure*}[t]
  \includegraphics[width=\linewidth]{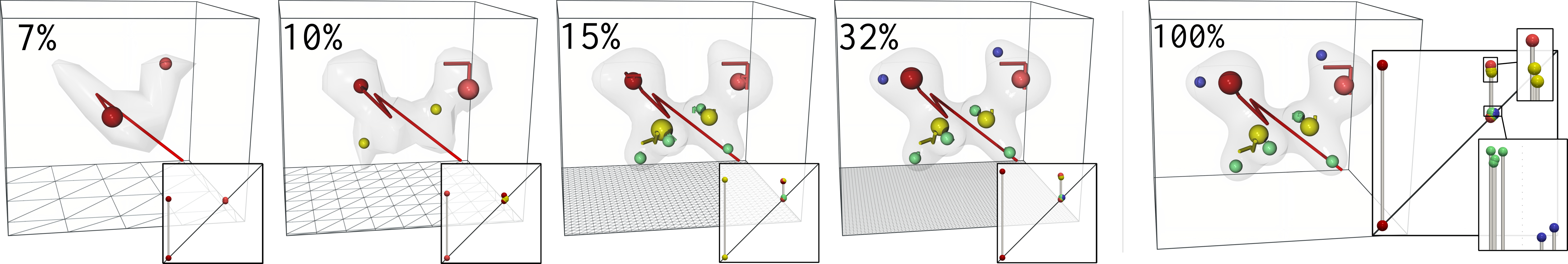}
\mycaption{Progressive 
persistence 
diagrams (saddle-maximum pairs, left to right) for the electron density of 
the ethane-diol molecule (transparent  isosurface), at a few steps of the 
progressive computation. Maxima (denoting the atoms) are shown in the domain 
with spheres, scaled by  persistence and \julien{colored by lifetime 
(red to blue)}, while their trajectory through the data hierarchy 
(\autoref{sec_cpLifeTime}) 
is shown with a curve 
(matching color). 
Our progressive approach captures the heaviest atoms first: two oxygens (at 
$7\%$ of the computation time), then two carbons ($10\%$) and finally the 
six hydrogens ($32\%$). 
}
\label{fig_ethaneDiol}
\end{figure*}

\section{Progressive Persistence Diagrams}
\label{sec_progressivePersistenceDiagram}


Our approach for progressive persistence diagrams 
leverages and 
combines the insights and algorithms introduced in the previous sections. It 
\julesReplace{initializes}{starts} at the coarsest hierarchy level, $\domain^{0}$, and then iterates 
progressively 
through the hierarchy levels,
producing
the exact 
persistence diagram $\diagram(f^{i})$ for each level $i$, until $i 
= h$.
We first introduce our approach 
in the non-progressive case 
(\autoref{sec_nonProgressive_diagram}, 
\autoref{fig_pd_nonProgressiveStrategy}), 
and then present our progressive strategy (\autoref{sec_progressiveDiagrams}). 
We focus on minimum-saddle persistence pairs, saddle-maximum pairs 
being treated symmetrically.



\subsection{Persistence Diagram from Critical Points}
\label{sec_nonProgressive_diagram}
The diagram $\diagram(f)$ 
\julienRevision{of the extremum-saddle pairs}
of an input field $f : \domain \rightarrow 
\range$ is computed as follows.
\julienReplaceMinor{}{In short, critical points are 
used as seeds for the computation of monotonic paths, specifically linking 
saddles down to minima. This first step identifies merge events occurring at 
saddle points (\emph{part A}). The merge events are processed in a second step 
(\emph{part B}) to track the connected components of sub-level sets. Similarly 
to previous topological techniques based on monotonic paths \cite{ChiangLLR05, 
MaadasamyDN12, CarrWSA16, smirnov17}, our approach emulates the usage of a 
Union-Find data structure with path compression \cite{cormen} (traditionally 
used for connectivity 
tracking) by propagating \emph{representants} between merged components. 
However our strategy is specialized for the production of persistence diagrams, 
and only visits monotonic paths of minimal length (i.e. integral lines).}


\noindent
\emph{Part A:}

\noindent
\emph{From data to reduced topological information}

\noindent
\textbf{1) Critical points.} First, critical points are extracted  
(\autoref{sec_criticalPoints}).

\noindent
\textbf{2) Saddle monotonic paths.}
The second step consists in initiating monotonic paths from each saddle $s$ 
downwards, to identify at least one minimum for each connected component of 
sub-level set merging at $s$ (\autoref{fig_pd_nonProgressiveStrategy}). 
For 
this,
we initiate 
backward integral lines (\autoref{sec_criticalPoints}), for each connected 
component of lower link $\lowerlink(s)$ of each saddle $s$. These integral 
lines are guaranteed to terminate in local minima of $f$. 
\julesReplaceMinor{In practice, o}{O}nce a backward integral line 
$\backwardIntegralLine(s)$ terminates in a local minimum $m$, we back-propagate 
the vertex identifier of $m$ and store it for each vertex $v \in 
\backwardIntegralLine(s)$. Then, $m$ is called 
a 
\emph{representant} of $v$, 
which is noted $r(v) = \{m\}$. This strategy enables the early termination of 
an integral line $\backwardIntegralLine(s_1)$ when it merges with another one, 
$\backwardIntegralLine(s_0)$, computed previously. In that case, we 
back-propagate the representants reported by the merge vertex on 
$\backwardIntegralLine(s_1)$ back to $s_1$. 
At the end of this step, each saddle $s$ is associated with the list of 
representants collected by its backward integral lines. These 
denote local minima 
which 
may 
have initially created the sub-level set 
components merging at $s$.


~

\noindent
\emph{Part B:}

\noindent
\emph{From reduced topological information to persistence diagrams}

\noindent
\textbf{3) Critical triplets.} For each saddle $s$, we create a list of 
\emph{critical triplets}, in the form $(s, m_0, m_1)$\julesRevision{, where
  $m_0$ and $m_1$ are representants of $s$ and thus are
local minima}.
These are obtained by 
considering \julesEditsOut{all the possible (order-independent) }pairs\julesEditsOut{,} among the set of 
representants of $s$ (computed previously).
Note that in 
practice, for nearly all saddles, this list \julesReplaceMinor{counts}{consists of} only one 
triplet, which describes the fact that $s$ separates two pits, $m_0$ and $m_1$. 
\julesEditsOut{Only in case of forking or degenerate saddles, 
multiple triplets can emerge at a given saddle.}
\julesRevision{\julienReplace{Note that}{However} in case of degenerate 
saddles, multiple triplets emerge.
For a degenerate saddle associated with $d$ representants $(m_0,\ldots,m_{d-1})$
in ascending values of $f$,
we create the $d-1$ triplets $(s,m_0,m_i)$ with $0<i<d$.}


\noindent
\textbf{4) Critical point pairing.} This step iterates over the global list of 
critical triplets (computed previously) in increasing order of saddle values.
The first triplet $(s_0, m_0, m_1)$ \julesReplaceMinor{models}{represents} the earliest merge 
event between connected components of sub-level sets of $f$. 
We introduce its \emph{simplified version}, $\big(s_0, r(m_0), 
r(m_1)\big)$, which is initially equal to $(s_0, m_0, m_1)$ (initially, a local 
minimum is itself its own representant).
The highest of the two minima, for instance $m_1$, is then selected to create 
in $\diagram(f)$ the critical point pair $(s_0, m_1)$. Indeed, since $s_0$ 
is the earliest merge event, $m_1$ is guaranteed to be the \emph{youngest} 
minimum, according to the Elder rule \cite{edelsbrunner09}, which created a 
component of sub-level set merging with another one at $s_0$. To model the 
\emph{death} of 
$m_1$'s
component (its merge with the component containing 
$m_0$), we 
update its representant 
as follows: $r(m_1) \leftarrow r(m_0)$. Thus, 
all future merging events involving $m_1$ will re-direct to $m_0$, as the 
component born at $m_1$ died by merging with that containing $m_0$ (following 
the Elder rule \cite{edelsbrunner09}). This simplification process is iterated 
over the (sorted) global list of critical triplets. 
At each step, when constructing a simplified triplet $\big(s, r(m_0), 
r(m_1)\big)$, we recursively retrieve the representants of $r(m_0)$ and 
$r(m_1)$, until we reach minima only representing themselves.
This guarantees that for 
each merge event of the sub-level set occurring at a saddle $s$, we can 
efficiently retrieve the deepest minimum for each of the 
components merging in $s$ and therefore pair it adequately in $\diagram(f)$.
\julesReplaceMinor{Note that this approach is equivalent to the use of 
union-finds to keep track of merge events at saddles 
points \julienMajorRevision{(in particular, the recursive update of 
representants is equivalent to the so-called \emph{path compression} of UF data 
structures \cite{cormen}). However, our approach only needs to process
integral 
lines (and not the entire triangulation).}
Note that our strategy
bears global
similarities with previous topological techniques based on 
monotonic paths
\cite{ChiangLLR05, MaadasamyDN12, CarrWSA16, smirnov17}, 
which in fact also use integral lines
as they provide the shortest monotonic paths. 
However, our strategy is specialized for the production of persistence diagrams.}{Note that the recursive update of 
representants is equivalent to the so-called \emph{path compression} of UF data 
structures \cite{cormen}.}
Overall, iterating as described above over the list of triplets results in 
populating $\diagram(f)$ with pairs from bottom to top (by increasing death 
values).

\subsection{Progressive Strategy}
\label{sec_progressiveDiagrams}
The above algorithm 
is divided in two parts 
(\emph{A} and 
\emph{B}, \autoref{sec_nonProgressive_diagram}). 
In particular, 
only part \emph{A} can leverage our progressive representation of the 
input data (\autoref{sec_progressiveData}), as 
part \emph{B} processes reduced topological information which \julesReplace{have}{has} been 
abstracted from it and which therefore become completely independent.
Thus, we focus our 
progressive strategy on part \emph{A}.
This has a negligible impact on practical \julesReplace{performances}{performance}. 
In \julesReplaceMinor{practice}{our experience}, 
part \emph{B} represents 
less than \jules{5\%} of the computation on average.
Critical points (Step 1) can be extracted progressively as described in 
\autoref{sec_progressiveCriticalPoints}. For Step 2, we investigated
multiple shortcut mechanisms (similar to \autoref{sec_cpRegularity}), to 
maintain the monotonic paths which remain valid from level $i$ to $i+1$.
However, from our experience, the overhead induced by this global maintenance 
is not compensated by the acceleration it induces at level $i+1$, as monotonic 
paths are usually highly localized and thus 
already inexpensive to compute (less than $10\%$ of the 
non-progressive computation on average).
Thus, our overall strategy for progressive persistence diagrams
simply consists, 
at 
each level $i$ of the triangulation hierarchy $\hierarchy$, in 
updating progressively the critical points 
(\autoref{sec_progressiveCriticalPoints}) 
and then triggering the fast, remaining steps of persistence diagram 
computation (2, 
3, 4) as described in \autoref{sec_nonProgressive_diagram}.

\subsection{Parallelism}
\label{sec_parallelPDalgortihm}
Our progressive algorithm for persistence diagram computation can be easily 
parallelized. The initial critical point computation (Step 1, 
\autoref{sec_nonProgressive_diagram}) is parallelized as 
described in \autoref{sec_cpParallel}. Saddle integration (Step 2, 
\autoref{sec_nonProgressive_diagram}) can be trivially parallelized over 
saddles. However, locks need to be used during representant back 
propagation (to guarantee consistency over concurrent accesses by distinct 
monotonic paths).
Critical triplet generation (Step 3, \autoref{sec_nonProgressive_diagram}) is 
also 
parallelized over saddles. In Step 4 (critical point pairing 
\autoref{sec_nonProgressive_diagram}), triplets are sorted in parallel using 
\julesReplaceMinor{GNU's efficent}{the efficient GNU} implementation \cite{singler2008gnu}. The reminder of Step 4 is 
intrinsically sequential (as representants need to be updated in order of 
simplification), but in practice, this step represents less than \jules{1\%} 
of the sequential execution, which does not impact parallel efficiency.

%

\begin{figure}
  \includegraphics[width=\columnwidth]{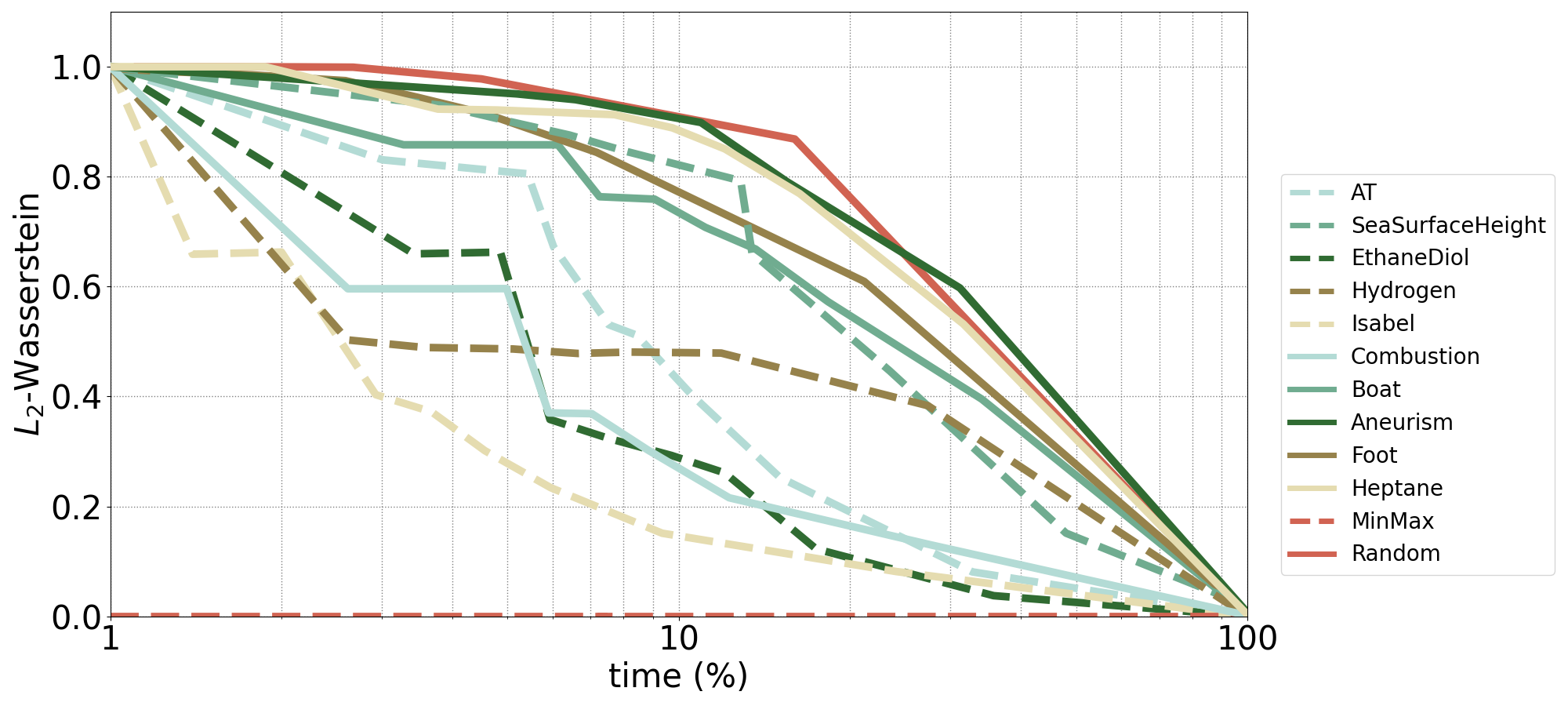}
  \caption{Empirical convergence of the normalized $L_2$-Wasserstein distance.
  Each curve plots the distance between the currently estimated diagram, 
  $\diagram(f^{i})$, and the final, exact diagram, $\diagram(f)$, as a function 
  of the percentage of computation time (logarithmic scale).
}
  \label{fig_convergence}
\end{figure}

\begin{figure}
  \includegraphics[width=\columnwidth]{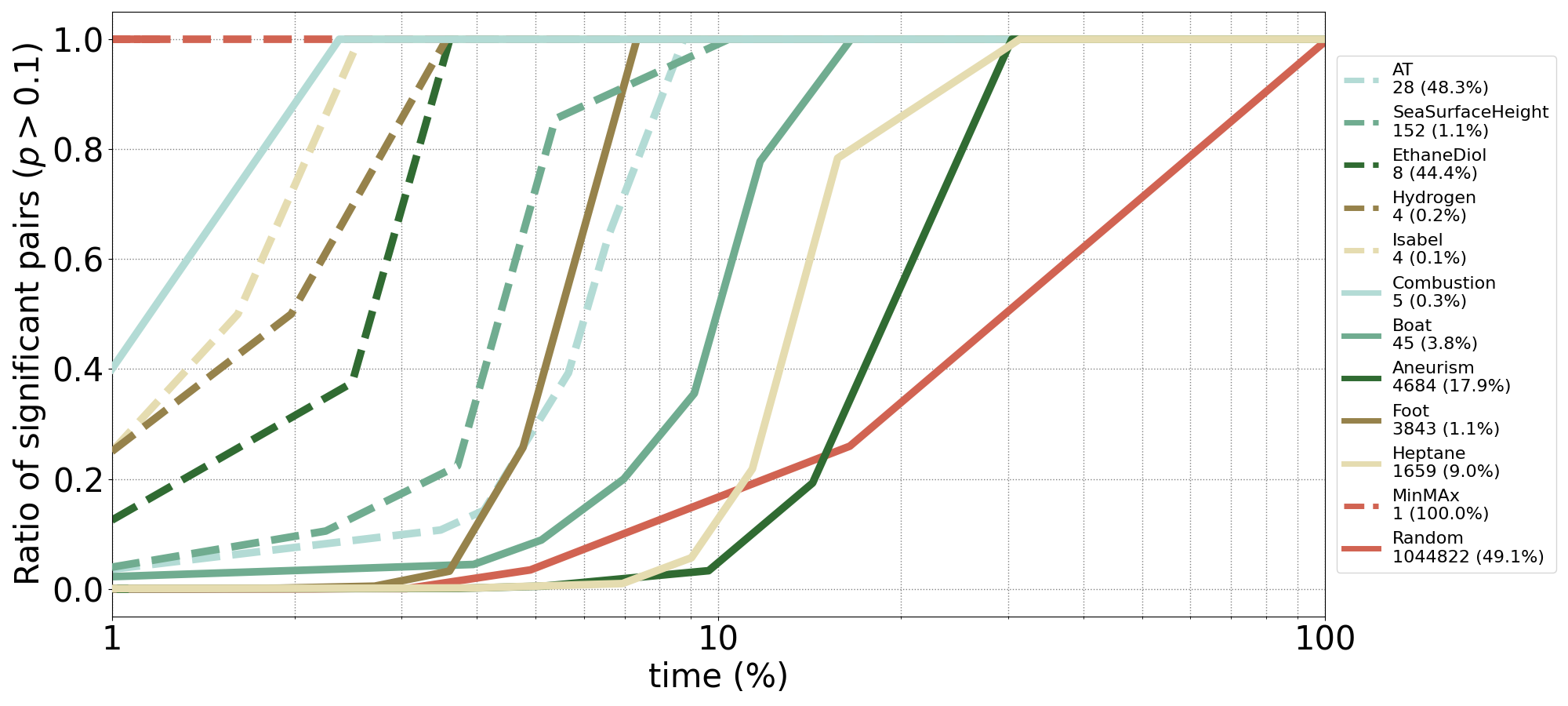}
  \caption{\julienRevision{Ratio of 
  captured 
  \emph{significant pairs} in 
$\diagram(f^i)$ (cf. \autoref{sec_result_data})
as a function of  computation time 
(logarithmic scale).
}}
  \label{fig_ratio_topPairs}
\end{figure}
\begin{figure}
  \includegraphics[width=\columnwidth]{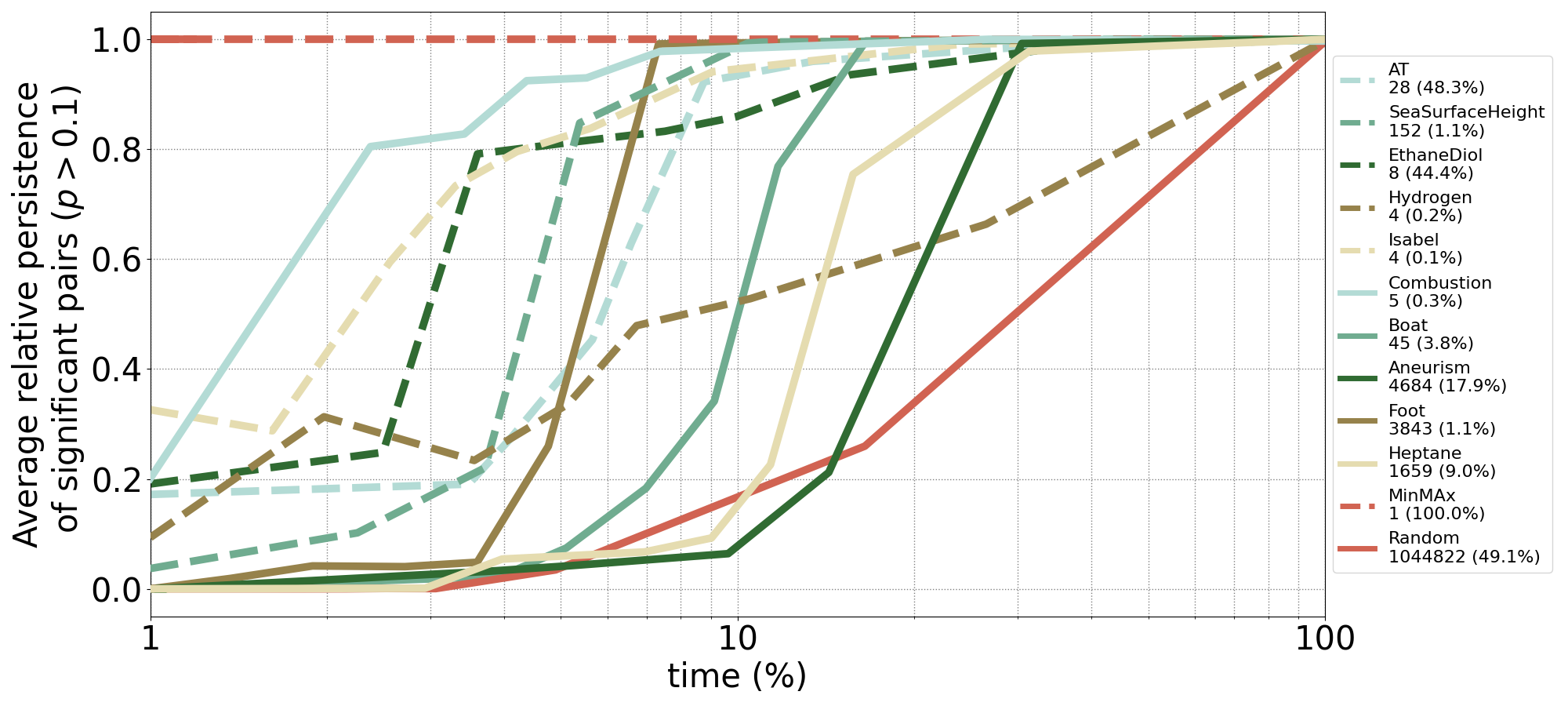}
  \caption{\julienRevision{Average persistence of the \emph{significant 
pairs} 
captured 
in $\diagram(f^i)$ 
(cf. \autoref{sec_result_data}),
relatively to the same
average
in $\diagram(f)$,
as a function of  computation time 
(logarithmic scale).}}
  \label{fig_average_topPairs}
\end{figure}
%

\section{Results}
\label{sec_results}
This section presents experimental results obtained on a 
computer with two Xeon CPUs (3.0 GHz, 2x4 cores, 64GB of RAM), with a C++ 
implementation of our algorithms \julienRevision{(publicly available at: 
\href{https://github.com/julesvidal/progressive-scalar-topology}{
https://github.com/julesvidal/progressive-scalar-topology})}, 
\julienRevision{written as}
modules for the 
Topology 
ToolKit (TTK) \cite{ttk17}.
The 
datasets 
\julienMajorRevision{are 3-dimensional (at the exception of 
\emph{SeaSurfaceHeight}, which is 2-dimensional) and they}
have been downloaded from public repositories 
\cite{openSciVisDataSets, ttkData}.

\subsection{Progressive Data Representation}
\label{sec_result_data}

In this section, we study 
the practical relevance of our progressive data 
representation (\autoref{sec_progressiveData}). 
First,
we 
evaluate its qualitative relevance. Our approach for persistence 
diagram computation (\autoref{sec_progressivePersistenceDiagram}) progressively 
refines an estimation of the  output $\diagram(f)$, by efficiently 
updating $\diagram(f^{i})$ at each new hierarchy level $i$. 
\julesEditsOut{To evaluate the 
quality of this estimation, for each level $i$, we measure 
the $L_2$-Wasserstein distance 
$\wasserstein{2}\big(\diagram(f), \diagram(f^{i})\big)$ 
(\autoref{sec_persistenceDiagram}).}
\julesRevision{To evaluate quantitatively the relevance of this estimation 
\julienMajorRevision{$\diagram(f^{i})\big)$}, we
  measure its \julienReplace{similarity}{distance}
to the final, exact result \julienMajorRevision{$\diagram(f)$} with the 
\emph{Wasserstein distance}, an established 
practical metric 
inspired by optimal 
transport \cite{Kantorovich, monge81}.
\julesReplaceMinor{}{Intuitively, this distance aims at optimizing a matching 
between the features of two diagrams  to compare and penalizes 
\emph{mismatches} between these diagrams.}
Given two diagrams $\diagram(f)$ and
$\diagram(g)$, a pointwise distance 
 $\pointMetric{q}$, inspired from the $L^p$ norm, can be introduced 
in the 2D birth/death space
between 
  two points $a = (x_a, y_a) \in \diagram(f)$ and 
$b = (x_b, y_b) \in \diagram(g)$, with $q > 0$, as: 
\vspace{-1.5ex}
\begin{equation}
\pointMetric{q}(a,b)=\left(|x_b-x_a|^q + |y_b-y_a|^q\right)^{1/q} = \|a-b\|_q
\label{eq_pointWise_metric}
\end{equation}
\vspace{-3.5ex}}

\noindent
\julesRevision{By convention, $\pointMetric{q}(a, b)$ is set to zero 
if both $a$ and $b$ exactly lie on the diagonal ($x_a = y_a$ and $x_b = y_b$).
The $L_q$-Wasserstein distance, noted 
$\wasserstein{q}$, between $\diagram(f)$ and 
$\diagram(g)$ can then be introduced as:
\begin{equation}
    \wasserstein{q}\big(\diagram(f), \diagram(g)\big) = 
\min_{\phi
\in \Phi} \left(\sum_{a \in \diagram(f)} 
\pointMetric{q}\big(a,\phi(a)\big)^q\right)^{1/q}
\label{eq_wasserstein}
\end{equation}
}

\noindent
where $\Phi$ is the set of all possible assignments $\phi$ mapping each 
point
$a \in \diagram(f)$ to 
a point
$b 
\in \diagram(g)$
or to 
its projection onto the diagonal.
$\wasserstein{q}$ can be computed 
via
assignment optimization, for which 
exact \cite{Munkres1957} and approximate \cite{Bertsekas81, Kerber2016}
implementations are publicly available \cite{ttk17}.
\julesReplaceMinor{Intuitively, this distance aims at optimizing a matching 
between the features of two diagrams  to compare ($\diagram(f)$ and
$\diagram(g)$) and penalizes 
\emph{mismatches} between these diagrams.}{}

\julesRevision{For each level $i$, we measure 
the $L_2$-Wasserstein distance 
$\wasserstein{2}\big(\diagram(f), \diagram(f^{i})\big)$.}
We normalize this distance by dividing it by 
$\wasserstein{2}\big(\diagram(f), \emptyset)$.
Then, along the hierarchy $\hierarchy$,
this normalized distance \julesReplaceMinor{evolves}{progresses} from $1$ to $0$ 
for all datasets. 
Although this distance may increase in theory from one level to the next, 
\julesReplaceMinor{Figure \ref{fig_convergence}}{\autoref{fig_convergence}} shows that
it is monotonically 
decreasing
for 
\julienReplace{our datasets (see the appendix for similar convergence 
curves on an extended collection of datasets, including two examples 
containing a minor oscillation at the beginning of the computation).}{all our 
datasets.} 
This shows that in practice, the accuracy of 
our progressive outputs
indeed improves over time.
\julienRevision{This empirical convergence evaluation gives a global picture of 
the quality of our progressive data representation. To further evaluate its 
relevance, we report in \autoref{fig_ratio_topPairs} 
the ratio of captured \emph{significant pairs} in the diagram $\diagram(f^{i})$ 
as a 
function of the computation time. 
To evaluate this ratio, we select the \emph{significant} pairs of  
$\diagram(f)$, i.e. with a relative persistence greater than $0.1$. Let $n_p$ 
be the number of such significant pairs (reported for each dataset in the 
legend of \autoref{fig_ratio_topPairs}, right, along with its percentage over 
the total number of pairs in $\diagram(f)$, in parenthesis).
Next, we select the $n_p$ 
most persistent pairs in $\diagram(f^{i})$ and divide the resulting number of 
selected pairs, noted $n_p^i \leq n_p$, by $n_p$.
In short, 
this indicator helps appreciate the number of significant features captured by 
the hierarchy \julesEditsOut{in its early levels}\julesRevision{early in the computation}. In particular, 
\autoref{fig_ratio_topPairs} shows that for most of the datasets, 
the number of captured significant pairs matches the final estimation as of 
$10\%$ of the computation time.
%
\julesReplaceMinor{Figure~\ref{fig_average_topPairs}}{\autoref{fig_average_topPairs}} reports the average persistence of the 
$n_p^i$ significant 
pairs in $\diagram(f^{i})$ as a 
function of the computation time, relatively to the
average persistence of the $n_p$ significant pairs in $\diagram(f)$. 
This indicator helps appreciate how well the 
significant
pairs are captured in the data hierarchy. In particular, this figure shows a 
clear global trend across datasets: the persistence of the significant pairs 
tends to 
be underestimated \julesEditsOut{in the early hierarchy levels}\julesRevision{early in the computation} and this estimation improves 
over time. 
These quantitative observations (early capture of the significant pairs and 
underestimation of persistence \julesEditsOut{at the early levels}\julesRevision{at the beginning of the computation}) can be visually observed in 
\autoref{fig_ethaneDiol}, which shows that the significant pairs 
are 
captured early in the data hierarchy (red and yellow pairs) but that their 
persistence is indeed underestimated: the corresponding points are 
initially close to the diagonal in the corresponding diagrams and then, they
progressively move away from it.}

Next, we evaluate 
the computational relevance of our 
progressive data representation, by 
reporting the number 
of Topologically Invariant (TI) vertices (\autoref{sec_topologicalInvariants}), 
for which no computation
is needed. 
Table \ref{tab_stats}
shows that for real-world datasets, 
TI vertices represent 
72\% of the data on average, which indicates that efficient update mechanisms 
can indeed be derived from
our progressive data representation.
\julienRevision{This table also includes the memory overhead induced in 
progressive mode by the data structures employed by our topological analysis 
algorithms (\autoref{sec_criticalPoints} and 
\autoref{sec_persistenceDiagram})\julienReplace{. In particular, this 
overhead is estimated by measuring the memory footprint of all the 
data-structures which are present in our progressive algorithms \julesReplaceMinor{(sections 
\ref{sec_progressiveCriticalPoints} and
\ref{sec_progressivePersistenceDiagram})}{(\autoref{sec_progressiveCriticalPoints}
and \autoref{sec_progressivePersistenceDiagram})}
but \emph{not} present in 
the TTK implementation of the state-of-the-art methods. Thus, this column 
depicts the additional 
memory needed by our approach in comparison to the standard procedures 
available in TTK. In particular, this column 
shows 
a linear evolution of this memory overhead with the size of the data 
hierarchy.}{, 
which exhibits a linear evolution with the size of the data hierarchy. }
Note 
that our implementation is not optimized for memory usage and that important 
gains can be expected by re-engineering our data structures at a low 
level.
}


%



\begin{table}
  \centering
    \rowcolors{3}{gray!10}{white}
  \resizebox{0.95\columnwidth}{!}{
  \begin{tabular}{|l||rrr||rr|r|r|}
\hline
{Dataset} &  $\sum_{i = 0}^{i = h} |\domain_0^i|$ & $h$ & M. (Mb)&\# old TIs & 
\# new TIs & Total TIs & \% \textbf{TIs} \\
\hline
AT                  & 931,110     & 9   & 242    & 93,145        & 509,504       & 602,649       & \textbf{64.7\%}  \\
SeaSurfaceHeight \julesRevision{\scriptsize{(2D)}}    & 1,384,626   & 11  & 240    & 241,264       & 608,460       & 849,724       & \textbf{61.4\%}  \\
EthaneDiol          & 2,057,388   & 9   & 527    & 227,376       & 1,371,537     & 1,598,913     & \textbf{77.7\%}  \\
Hydrogen            & 2,413,532   & 8   & 626    & 245,541       & 1,528,607     & 1,774,148     & \textbf{73.5\%}  \\
Isabel              & 3,605,604   & 9   & 970    & 274,590       & 1,329,116     & 1,603,706     & \textbf{44.5\%}  \\
Combustion          & 4,378,386   & 9   & 1,149  & 421,208       & 2,445,192     & 2,866,400     & \textbf{65.5\%}  \\
Boat                & 4,821,326   & 9   & 1,221  & 575,646       & 3,690,624     & 4,266,270     & \textbf{88.5\%}  \\
Random              & 18,117,518  & 9   & 5,389  & 169,603       & 54            & 169,657       & \textbf{0.9\%}   \\
MinMax              & 18,994,899  & 9   & 4,742  & 2,394,619     & 16,474,429    & 18,869,048    & \textbf{99.3\%}  \\
Aneurism            & 19,240,277  & 9   & 4,841  & 2,367,190     & 16,027,209    & 18,394,399    & \textbf{95.6\%}  \\
Foot                & 19,240,277  & 9   & 5,109  & 1,648,823     & 10,746,624    & 12,395,447    & \textbf{64.4\%}  \\
Heptane             & 31,580,914  & 10  & 8,117  & 3,453,180     & 22,512,477    & 25,965,657    & \textbf{82.2\%}  \\
\hline
\end{tabular}

  }
  \vspace{0.5em}
\caption{
\julienRevision{Statistics of our progressive data hierarchy. From left to 
right: number of vertices, number of levels, memory 
\julienReplace{overhead (over TTK)}{footprint}, and number}
of topologically invariant (TI) vertices 
(\autoref{sec_topologicalInvariants}) in the data hierarchy.
For real-world datasets (Random and MinMax excluded), topologically invariant 
vertices represent 72\% of the data on average.}
%
%
 \label{tab_stats}
\end{table}

\begin{table}[t]
\vspace{-1ex}
    \rowcolors{3}{gray!10}{white}
    \centering
    \resizebox{0.95\columnwidth}{!}{
    \begin{tabular}{|l|r|rrr|r| rrr|}
\hline
    &\multicolumn{4}{c|}{Critical Points}&\multicolumn{4}{c|}{Persistence Diagram}\\
 Dataset &TTK \cite{banchoff70}&NP&Prog &{\small Speedup}& TTK 
\cite{gueunet_tpds19}& NP &  Prog & {\small Speedup}\\
\hline
AT & 4.41 & 0.34 & 0.25 & 1.36 & 0.66 & 0.31 & 0.24 & 1.29\\
SeaSurfaceHeight \julesRevision{\scriptsize{(2D)}}& 0.92 & 0.16 & 0.26 & 0.62 & 0.70 & 0.24 & 0.40 & 0.60\\
EthaneDiol & 9.56 & 0.73 & 0.42 & 1.74 & 1.45 & 0.68 & 0.41 & 1.66\\
Hydrogen & 11.55 & 0.92 & 0.59 & 1.56 & 1.90 & 0.89 & 0.63 & 1.41\\
Isabel & 17.98 & 1.40 & 1.43 & 0.98 & 2.76 & 1.50 & 1.62 & 0.93\\
Combustion & 21.87 & 1.74 & 1.33 & 1.31 & 4.36 & 1.82 & 1.50 & 1.21\\
Boat & 22.47 & 1.74 & 0.73 & 2.38 & 3.38 & 1.81 & 0.82 & 2.21\\
Random & 113.12 & 13.99 & 21.04 & 0.66 & 74.39 & 25.65 & 34.40 & 0.75\\
MinMax & 82.23 & 6.94 & 1.56 & 4.45 & 14.68 & 7.00 & 1.64 & 4.27\\
Aneurism & 84.03 & 7.39 & 2.19 & 3.37 & 12.85 & 8.03 & 3.43 & 2.34\\
Foot & 100.58 & 9.26 & 8.13 & 1.14 & 18.20 & 12.18 & 12.06 & 1.01\\
Heptane & 149.30 & 12.43 & 6.46 & 1.92 & 19.45 & 13.22 & 8.16 & 1.62\\
\hline
\end{tabular}

  }
  \vspace{0.5em}
    \caption{\julienRevision{Sequential computation times (in seconds)} of our 
algorithms 
for critical point extraction (left) and persistence diagram computation 
(right). The columns \emph{TTK} report the run times of the default 
implementations provided by the Topology ToolKit \cite{ttk17}. The columns 
\emph{NP} and \emph{Prog} respectively report the timings for the 
non-progressive (directly initialized at the final hierarchy level)  and  
progressive versions of our algorithms.}
  \label{tab_sequentialTimings}
\end{table}
\begin{table}
    \rowcolors{3}{gray!10}{white}
    \centering
    \resizebox{0.95\linewidth}{!}{
    \begin{tabular}{|l|r|rrr|r|rrr|}
\hline
    &\multicolumn{4}{c|}{Critical Points}&\multicolumn{4}{c|}{Persistence Diagram}\\
 Dataset &TTK \cite{banchoff70}&NP&Prog &{\small Speedup}& TTK 
\cite{gueunet_tpds19}& NP &  Prog & {\small Speedup}\\
\hline
AT & 0.54 & 0.06 & 0.05 & 1.20 & 0.29 & 0.06 & 0.06 & 1.00\\
SeaSurfaceHeight \julesRevision{\scriptsize{(2D)}}& 0.15 & 0.04 & 0.05 & 0.80 & 0.28 & 0.07 & 0.11 & 0.64\\
EthaneDiol & 1.18 & 0.12 & 0.07 & 1.71 & 0.42 & 0.13 & 0.11 & 1.18\\
Hydrogen & 1.41 & 0.14 & 0.12 & 1.17 & 0.97 & 0.18 & 0.19 & 0.95\\
Isabel & 2.11 & 0.21 & 0.21 & 1.00 & 0.89 & 0.25 & 0.29 & 0.86\\
Combustion & 2.54 & 0.25 & 0.21 & 1.19 & 0.82 & 0.29 & 0.30 & 0.97\\
Boat & 2.71 & 0.26 & 0.15 & 1.73 & 0.81 & 0.30 & 0.24 & 1.25\\
Random & 11.46 & 1.82 & 2.67 & 0.68 & 19.10 & 8.96 & 10.89 & 0.82\\
MinMax & 9.87 & 1.25 & 0.43 & 2.91 & 3.24 & 1.39 & 0.69 & 2.01\\
Aneurism & 10.19 & 1.14 & 0.50 & 2.28 & 4.08 & 1.89 & 1.48 & 1.28\\
Foot & 11.30 & 1.93 & 1.75 & 1.10 & 5.39 & 3.37 & 3.29 & 1.02\\
Heptane & 17.72 & 2.38 & 1.51 & 1.58 & 5.74 & 2.79 & 2.50 & 1.12\\
\hline
\end{tabular}

    }
  \vspace{0.5em}
  \caption{\julienRevision{Parallel computation times (in seconds, 8 cores) of 
our 
algorithms 
for critical point extraction (left) and persistence diagram computation 
(right). The columns \emph{TTK} report the run times of the default 
implementations provided by the Topology ToolKit \cite{ttk17}. The columns 
\emph{NP} and \emph{Prog} respectively report the timings for the 
non-progressive (directly initialized at the final hierarchy level)  and  
progressive versions of our algorithms.}}
    \label{tab_parallel}
\end{table}

\begin{figure*}
  \includegraphics[width=\linewidth]{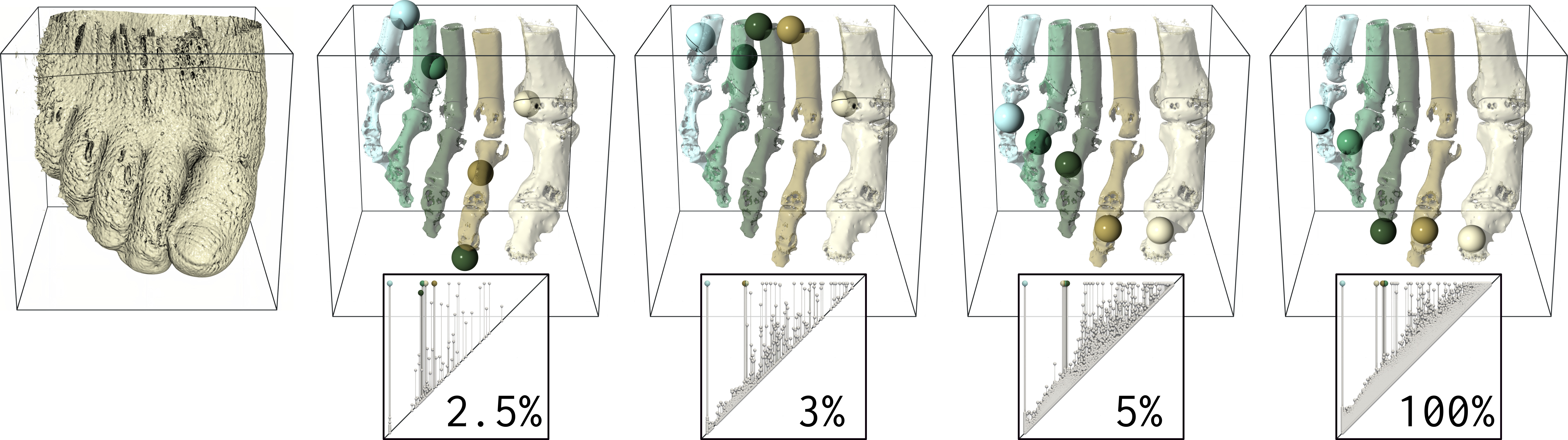}
  \mycaption{Progressive persistence diagrams (saddle-maximum pairs, from left 
to 
right) of the CT scan of a foot (leftmost: isosurface), at a few steps of the 
computation. 
A merge tree based segmentation (colored regions, computed 
with TTK \cite{ttk17}) reveals the $5$ most persistence structures in the data. 
Colored spheres show the
$5$ most persistent maxima reported by the current diagram estimation, 
illustrating a correct capture of the main structures early in the 
computation (as of $3\%$ of computation).}
\label{fig_foot}
\end{figure*}

\subsection{Time Performance}
\label{sec_resultsCriticalPoints}
The time complexity of our progressive 
algorithm for critical point extraction is linear 
with the number of input vertices,
which results in our hierarchical setup in 
$\mathcal{O}(\sum_{i = 0}^{i= h} |\domain_0^i|)$
steps. 
For
persistence diagrams, in the worst possible 
configuration (degenerate saddles 
with systematic integral line forking), each saddle would generate \julesRevision{monotonic}\julesEditsOut{monotone}
paths which would hit every minimum. This would yield 
\julesEditsOut{$n_s \times \big(n_m \times (n_m - 1)\big)/2$}
\julesRevision{$n_s \times (n_m - 1)$}
critical triplets, where $n_s$ and $n_m$ stand for the 
number of saddles and minima of $f$. 
\julesRevision{This would yield \julesReplaceMinor{for the critical pairing step }{}$n_s\times (n_m-1)$
  merge events \julesReplaceMinor{}{for the critical pairing step, each }with
  an amortized complexity \julesReplaceMinor{}{of
  }$\mathcal{O}\big(\alpha(n_m)\big)$, where $\alpha$
is the inverse of the Ackermann function.}
However, such configurations are extremely 
rare in practice and most saddles only yield one triplet, resulting in an 
overall practical time complexity of 
\julesEditsOut{$\mathcal{O}\big( \sum_{i = 0}^{i= h} (|\domain_1^i| + n_s^i)\big)$ steps.}
\julesRevision{$\mathcal{O}\big( \sum_{i = 0}^{i= h} (|\domain_1^i| + n_s^i\log n_s^i + n_s^i\alpha(n_m^i))\big)$ steps, 
also accounting for the sorting of triplets.}


Table \ref{tab_sequentialTimings} reports computation times (sequential run) 
for the default algorithms \julienRevision{(\autoref{sec_criticalPoints})}
\cite{banchoff70}, 
\cite{gueunet_tpds19} available 
in 
TTK \cite{ttk17} and the 
non-progressive and progressive versions of our algorithms. 
Non-progressive methods (\emph{TTK} and \emph{NP} columns) 
compute from scratch only the last hierarchy level $h$ directly.
We only report the run times of 
TTK as 
an indicative baseline as the differences in triangulation implementations
already induce alone important run 
time variations (TTK emulates implicitly triangulations for regular grids 
\emph{at query time}, while our implementation stores the explicit list of 
link edges for each vertex, \autoref{sec_cpUpdates}). 
Interestingly, 
the \emph{Speedup} columns 
show that, in addition to their ability to provide continuous 
visual feedback, our progressive algorithms 
are also
faster than 
their non-progressive versions
(on 
average, $1.8$ times faster for critical points, $1.6$ for persistence 
diagrams).
%
These 
speedups confirm that the overhead of processing an entire 
hierarchy 
($\sum_{i = 0}^{i = h} |\domain_0^i|$ vertices in progressive 
mode,
instead of  $|\domain_0^h|$ in 
non-progressive mode) and of detecting TI vertices is 
largely compensated by the gains these vertices provide.
Note that the datasets 
with
the most (resp. least) 
TI
vertices  (\autoref{tab_stats}) are also those for which the largest 
(resp. smallest) speedups are obtained, confirming the importance of 
TI
vertices in the computation.



Table \ref{tab_parallel} details the performance of the shared-memory 
parallelization of our progressive algorithms, using OpenMP 
\cite{dagum1998openmp}\julienRevision{, again in comparison to the default 
algorithms \julienRevision{(\autoref{sec_criticalPoints})} \cite{banchoff70}, 
\cite{gueunet_tpds19} available in 
TTK \cite{ttk17} and to the non-progressive version of our algorithms}. 
As mentioned in \autoref{sec_cpParallel}, critical point 
extraction can be trivially parallelized over vertices, for each of the four 
steps of our algorithm, resulting in an average parallel efficiency of 66\%. The 
persistence diagram computation results in a more modest efficiency (45\%) as 
monotonic path computations are subject to locks, in addition to be possibly 
imbalanced.

\subsection{\julienRevision{Stress Cases}}
\julienRevision{Our experiments include two synthetic datasets, whose purpose 
is to illustrate the most and the least favorable configurations for our 
approach, to better appreciate the dependence of our algorithms to their 
inputs. The \emph{MinMax} dataset is an elevation field which only contains 
one global minimum and one global maximum.
It exhibits therefore a lot of 
regularity. In contrast, the \emph{Random} dataset assigns a random value to 
each vertex. Thus, no local coherency can be expected between consecutive levels 
in the data hierarchy (which is an important hypothesis in our framework).}

\julienRevision{Table~\ref{tab_stats} confirms the best/worst case aspect 
of these datasets, as they respectively maximize and minimize the ratio of TI 
vertices: \emph{MinMax} has nearly only TI vertices ($99.3\%$) while 
\emph{Random} has nearly none ($0.9\%$).} 

\julienRevision{Table~\ref{tab_sequentialTimings} confirms, as can be expected, 
that these two datasets also maximize and minimize the speedup induced by our 
progressive approach. In particular, our progressive algorithms report a 
speedup greater than $4$ over their non-progressive versions
for the \emph{MinMax} dataset. This further confirms the 
observation made in \autoref{sec_resultsCriticalPoints} that processing an 
entire data hierarchy with the acceleration induced by TI vertices can indeed 
be faster than computing criticality from scratch at the final hierarchy level 
only (in particular, up to $4$ times). In contrast, this table also shows that 
in the worst possible case (\emph{Random}, nearly no TI vertices), the 
processing of the entire hierarchy can be up to $50\%$ slower (for critical 
points, $30\%$ for persistence diagrams) than computing in 
non-progressive mode at the final hierarchy level only. All the other datasets 
exhibit speedups included within these lower (\emph{Random}) and upper 
(\emph{MinMax}) bounds (on average $1.8$ for critical points, $1.6$ for 
persistence diagrams).}

\julienRevision{In terms of quality, the best/worst case aspect of 
\emph{MinMax} and \emph{Random} is also illustrated in 
Figs.~\ref{fig_convergence},~\ref{fig_ratio_topPairs} 
and~\ref{fig_average_topPairs}, where \emph{MinMax} converges 
immediately, while \emph{Random} describes the worst case (slow 
convergence, slow and inaccurate capture of the significant pairs). 
In these curves, the other datasets cover the span of possible behaviors 
between these two extreme cases.}

\subsection{Progressive Topological Visualization and Analysis}
\label{sec_resultsUI}

This section discusses the progressive visualizations and analyses enabled by 
our approach.
\julesReplaceMinor{Figure \ref{fig:teaser}}{\autoref{fig:teaser}} presents a typical example of progressive persistence 
diagram
computation 
on the electron density of the adenine-thymine (AT) molecular system. 
\julienRevision{In this figure, the}
estimated diagrams 
progressively capture the features in a meaningful way, as the heaviest atoms 
are captured first and the lighest ones last. In particular, in the diagrams, 
the introduced points 
progressively stand out from the diagonal towards their 
final locations.
As of $33\%$ of the computation, the diagram is complete and its 
accuracy is further improved over time.
This illustrates the capacity of our approach to deliver 
\julien{relevant
previews of the topological features of a dataset} and to improve them 
progressively. \julesReplaceMinor{Figure \ref{fig_ethaneDiol}}{\autoref{fig_ethaneDiol}} further illustrates our estimation of 
the lifetime of extrema and their trajectory in the data hierarchy. 
There, as one progresses down the hierarchy, the 
\julienRevision{prominent} maxima are \julienRevision{progressively} captured 
and they
quickly stabilize in 
the vicinity of their final location.
\julesReplaceMinor{Figure \ref{fig_foot}}{\autoref{fig_foot}} illustrates progressive persistence diagrams for an 
acquired dataset. There, a merge tree based segmentation (computed with TTK 
\cite{ttk17}) is shown in the background. 
It represents the regions of the five most persistent leaf arcs of the merge 
tree. The five most persistent maxima reported by the current diagram 
estimation are reported with spheres. As of $3\%$ of the computation, these 
maxima are correctly assigned to the final structures (one per toe), while their 
positional accuracy is further improved with time. Overall, the diagrams 
(bottom) capture the main features early in the computation, while smaller 
features and noise are progressively captured as the computation unfolds.
%
%
\begin{figure}[t]
  \centering
  \includegraphics[width=0.845\linewidth]{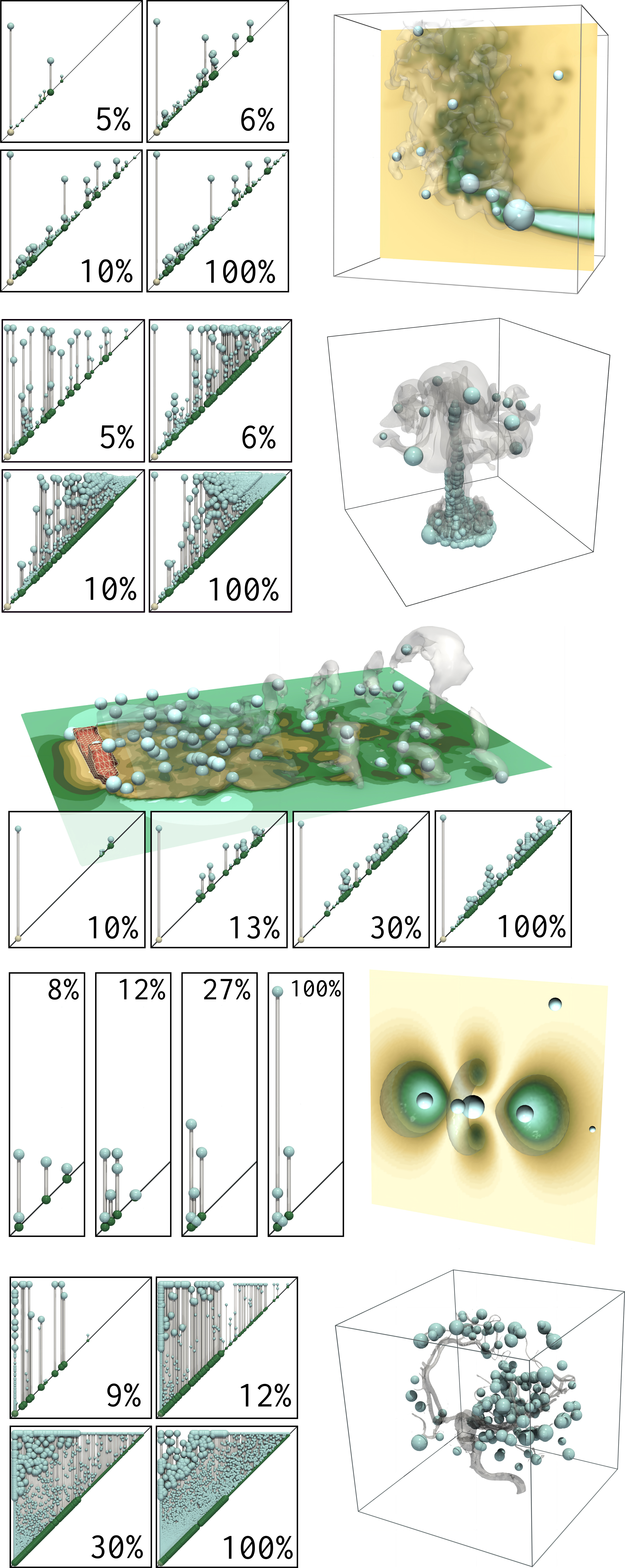}
  \mycaption{Progressive persistence diagrams (saddle-maximum pairs) 
for several data sets 
\julien{(combustion, heptane, boat, hydrogen, aneurism)},
at a 
few steps of the computation. 
Persistent maxima are represented with spheres in the 
domain (scaled by persistence). 
The progressive 
diagrams capture well the overall shape 
(number 
and salience of features)
of the final, exact output ($100\%$) early in the computation and  
refine it over time.
} 
\label{fig_gallery}
\end{figure}
\julesReplaceMinor{Figure \ref{fig_gallery}}{\autoref{fig_gallery}} presents a
gallery of progressive persistence diagrams 
for several datasets. 
The diagram 
estimations capture well the overall shape of the final, exact output 
(i.e. the number and salience of its main features)
and 
are progressively refined over time. This gallery complements the quantitative 
analysis reported in 
\julienRevision{Figs.~\ref{fig_convergence}, \ref{fig_ratio_topPairs}, 
\ref{fig_average_topPairs}}
and confirms visually the 
interest
of our progressive representations, which provide 
relevant
previews 
of 
the topological features present in a dataset.

\julien{We used the TTK library \cite{ttk17} to integrate our implementation 
within the popular visualization system ParaView \cite{paraviewBook}, as 
shown in the companion video (supplemental material).
This video illustrates the progressive updates of our topological 
previews within interactive times, and further demonstrates their interest for 
interactive visualization environments.}

\begin{figure}
  \centering
  \includegraphics[width=.95\columnwidth]{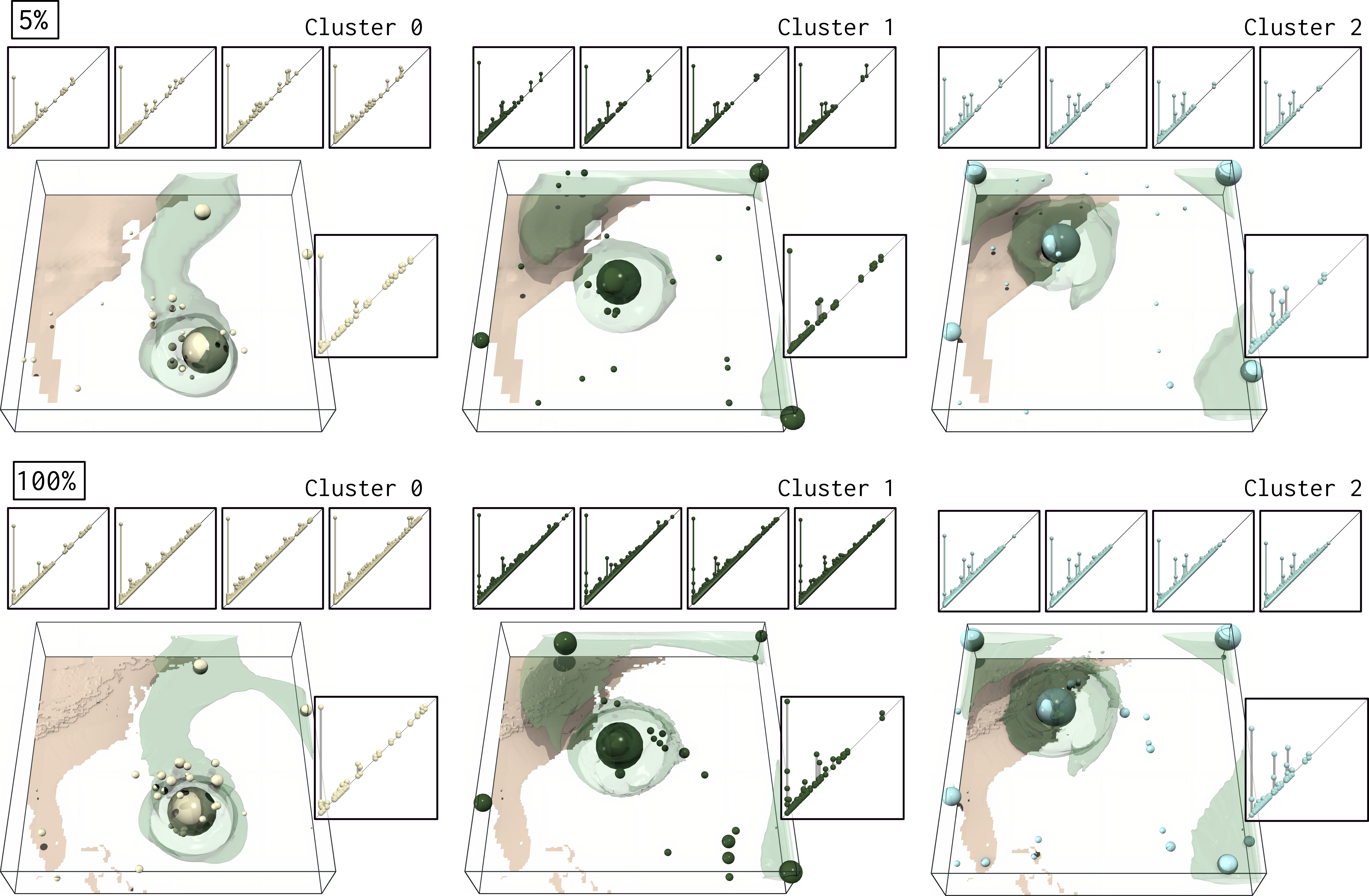}
  \mycaption{
  \julien{
  Topological clustering of the Isabel ensemble dataset.
  Progressive persistence diagrams (top, interrupted at \julien{$5\%$} 
  of 
  computation)
  are used as an input to the clustering approach by Vidal et al. 
\cite{vidal20} (constrained to $1$ second of computation). This 
time-constrained ensemble clustering yields the same, correct classification 
(one color per cluster, from left to right) as the one returned 
with the
exact diagrams (bottom).}}
\label{fig_clustering}
\end{figure}

\julesReplaceMinor{Figure \ref{fig_clustering}}{\autoref{fig_clustering}} illustrates the interest of 
our progressive representations for the control of the run time of 
batch-mode analysis pipelines. 
We consider the Isabel ensemble \cite{scivisIsabel, ttkData} (12 members 
illustrating 3 
hurricane behaviors: 
formation, drift and landfall, \autoref{fig_clustering}, left to right). 
Our progressive algorithm is used to generate a persistence diagram for each 
member, and is interrupted at a predefined threshold of computation time 
(\autoref{fig_clustering}, top). Then,
the TTK implementation of the algorithm by Vidal et al. \cite{vidal20} 
is used to cluster these diagrams (with a time constraint of 
one second). Overall, this results in a topological clustering pipeline whose 
time execution is 
fully 
controlled, from the feature extraction to their 
clustering. For reasonable 
computation thresholds 
(\julien{5\%} 
in 
\autoref{fig_clustering}), this pipeline returns 
the same, correct classification (one color per cluster) as
the one returned with the exact diagrams (bottom). This demonstrates that the 
main trends of an ensemble in terms of features (the main clusters) can still 
be estimated reliably\julesReplaceMinor{ in practice}{}, while
additionally controlling the execution time of the clustering pipeline.

\subsection{Limitations and Discussion}
\label{sec_limitations}
Our progressive persistence diagrams tend in practice to capture the 
main features first. However, this  
cannot be guaranteed  theoretically.
For instance, sharp spikes in the data \julienRevision{(e.g. high amplitude 
\emph{and} high frequency noise)}
 can yield persistent maxima 
only at the last levels of the hierarchy\julienRevision{, as illustrated in 
\autoref{fig_gallery} where the global maximum of the \emph{Hydrogen} dataset 
(fourth row)
belongs to a sharp spike in the center of the data (as also reported by the 
quantitative plots Figs.~\ref{fig_convergence} and 
~\ref{fig_average_topPairs})}.
This
behavior 
prevents the definition of theoretical error bounds on our estimations.
However, the empirical monotonic decrease of the Wasserstein 
distance 
(\autoref{fig_convergence}) 
indicates that our progressive representations \julesReplaceMinor{}{actually }provide reliable estimations\julesReplaceMinor{ in 
  practice}{}\julienRevision{, as confirmed by the indicators of
Figs.~\ref{fig_ratio_topPairs},~\ref{fig_average_topPairs}, where the 
real-world datasets  cover the span of possible behaviors between the 
two stress cases (\emph{MinMax}, \emph{Random}). This can be explained by 
the fact that, in practice, persistent pairs often coincide with large features 
in the domain, which get captured early in the data hierarchy.}
%
%
%
%

Although we described our approach generically, we focused in this paper on an 
efficient implementation of edge-nested triangulations for regular grids 
(\autoref{sec_gridHierarchy}). The generalization of our approach to 
generic domains requires to investigate triangulation subdivision schemes. 
Several of them seem compliant with the notion of
edge-nested triangulation (\autoref{sec_multiRes}), such as 
 the Loop subdivision \cite{loop87} 
 \julienRevision{and the \emph{red} triangulation refinement 
\cite{freudenthal42, bank83, bey95, zhang95}.}
However, efficiently 
transforming an arbitrary triangulation into a 
\julienRevision{triangulation which admits an edge-nested hierarchy}
is an orthogonal question which we leave for future work. Similarly, 
the reliable tracking of extrema through the hierarchy (for lifetime 
estimation, \autoref{sec_cpLifeTime}) relates to another orthogonal problem, for 
which computationally expensive optimizations 
may
need to be considered.
Our algorithms complete a given hierarchical level before moving on to 
the next one. \julesReplaceMinor{In practice, t}{T}his results in increasing update times as the 
computation converges. In the future, finer update strategies will be 
considered\julienRevision{, by considering adaptive, feature-centric, 
variable level-of-detail refinement methods}. Finally, our algorithm for 
persistence diagrams does not support 
saddle-saddle pairs in 3D. However, from our experience, the 
interpretation of these structures is not obvious in the applications.

Our progressive scheme 
seems to be particularly efficient 
for algorithms which visit \emph{all} the vertices of the domain 
(e.g. critical point extraction), but less beneficial for inexpensive 
operations which 
only visit small portions of the data 
(e.g.
integral line computation, \autoref{sec_progressiveDiagrams}). 
This is a 
lesson learned from our experiments which could serve 
as guideline for future 
extensions to other topological analyis algorithms.
\julesRevision{\julienReplace{Also, there}{There} is a trade off between the 
benefits of the 
progressive scheme and its cost in \julienReplace{terms}{term} of memory usage. 
Future work is needed
to improve the memory footprint of our approach by optimizing our data structures at a low level.
\julienReplace{For instance, for triangulations of regular grids and real-life 
tetrahedral meshes, the maximum number of neighbors around a vertex is typically 
small, 
which enables the encoding of local neighbor identifiers with very few 
bits, instead of full integers (as done in our current implementation). 
Other variables (such as the polarity, currently 
stored with a boolean for each 
neighbor) could also benefit from a more compact bit representation.}{For 
regular grids in particular, there are ways to take advantage of the low
number of local neighbors to store their indices in a lower number of bytes, or
use bit manipulation methods to encode all the informatin about a vertex in a 
small integer.}}

\section{Conclusion}
\label{sec_conclusion}

This paper introduced an approach for the progressive topological analysis of 
scalar data. Our work is based on a hierarchical representation of the input 
data and the fast identification of \emph{topologically invariant vertices}, 
for which we showed that no computation was required as they were introduced in 
the hierarchy. This enables the definition of efficient coarse-to-fine 
topological algorithms, capable of providing \julesReplace{exploitable}{interpretable} outputs upon 
interruption requests, and of progressively refining them otherwise until the 
final, exact output. We instantiated our approach with two examples of 
topological algorithms (critical point extraction and persistence 
diagram computation), 
which leverage efficient update mechanisms for ordinary vertices and avoid 
computation for the topologically invariant ones. 
For real-life datasets,
our algorithms tend to first capture the most important features of the 
data and to progressively refine their estimations with time. This is 
confirmed quantitatively with the empirical convergence of the Wasserstein 
distance to the final, exact output,
which is monotonically decreasing. More computation time indeed results in more 
accuracy. Our experiments also reveal that our progressive computations
even turn out to be faster overall than 
non-progressive algorithms and 
that they can be further accelerated with shared-memory parallelism.
We showed the interest of 
our approach 
for interactive data exploration, where our 
algorithms
provide 
progressive
previews, continuously refined over time, of the topological features 
found in a dataset.
We also showed that in batch-mode, our approach 
enables to 
control the run time of a complete TDA pipeline 
(topological clustering of ensemble data).

We believe our work sets 
the foundations for several exciting research 
avenues 
for future work.
First, we identified several improvement directions regarding the management of 
the input hierarchy, including the extensions to arbitrary triangulations 
or 
the addition of intermediate hierarchy levels.
\julienMajorRevision{Second, since our progressive algorithms can be given a 
time-budget constraint and still be resumed afterwards if needed, we would like 
to investigate in the future how such preemptable data analysis algorithms 
can help for optimizing scheduling policies in high-performance 
environments, where data analysis is often run at the same time as data 
production and where the allocation of computation resources need to be finely 
optimized.}
\julienReplace{Third}{Second}, we believe our approach can be generalized to 
higher dimensions (with 
tailored sampling methods) as well as to
other topological abstractions, in order to 
re-visit the entire TDA arsenal (merge trees, Reeb graphs, Morse-Smale 
complexes) in the light of progressivity. In that perspective, the 
generalization of topologically invariant vertices 
to 
Discrete Morse Theory \cite{forman98} looks particularly promising.
\ifCLASSOPTIONcompsoc
  \section*{Acknowledgments}
\else
\fi


We would like to thank the reviewers for their thoughtful remarks and
suggestions.
This work is partially supported by the European Commission grant 
H2020-FETHPC-2017 \emph{``VESTEC''} (ref. 800904).

\ifCLASSOPTIONcaptionsoff
  \newpage
\fi



%
\bibliography{paper}
\bibliographystyle{IEEEtran}



%
\vspace{-1ex}
\begin{IEEEbiography}[{\includegraphics[width=1in,height=1.25in,clip,
keepaspectratio]{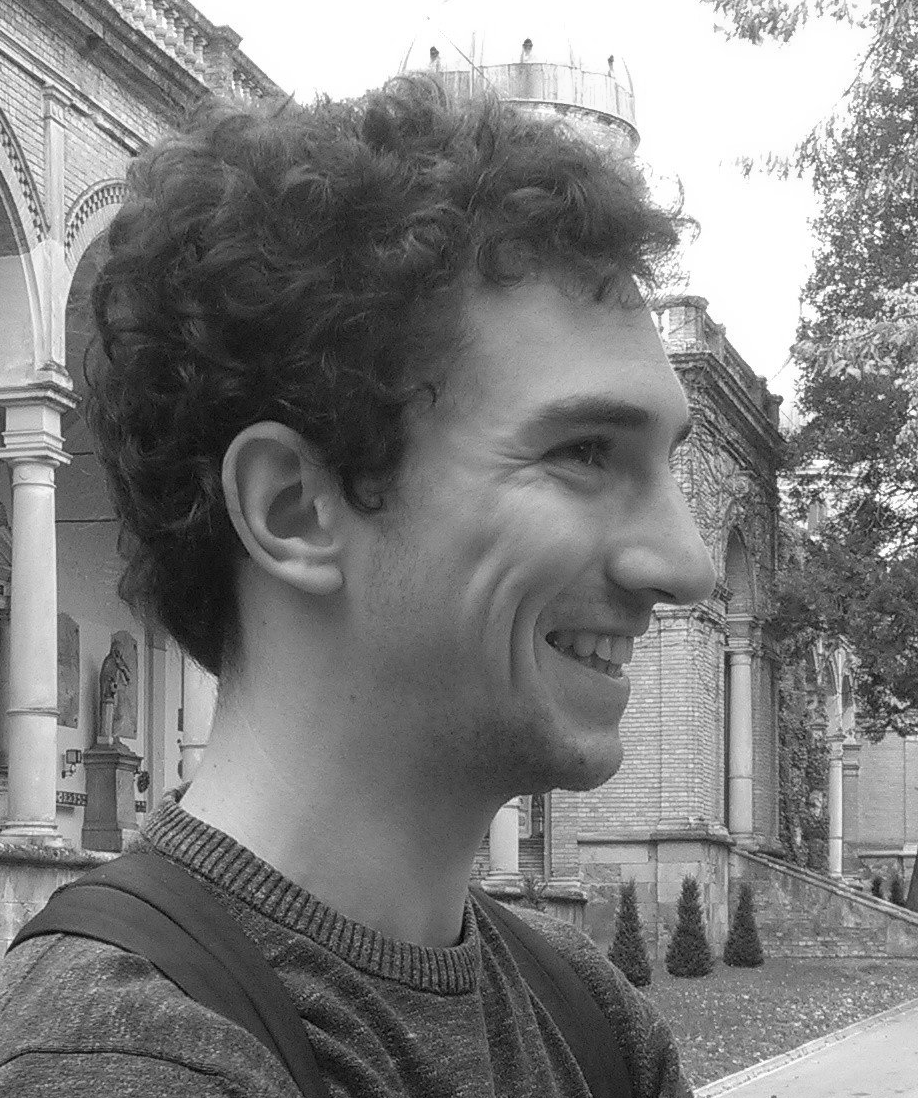}}]{Jules Vidal}
is a Ph.D. student at Sorbonne Université. He received the engineering degree
in 2018 from ENSTA Paris. He is an active contributor to 
the Topology ToolKit 
(TTK), an open source library for 
topological data analysis.
His notable contributions to TTK include the
efficient and progressive approximation of distances, barycenters and 
clusterings of persistence diagrams.
\end{IEEEbiography}

\vspace{-1ex}
\begin{IEEEbiography}[{\includegraphics[width=1in,height=1.25in,clip,
keepaspectratio]{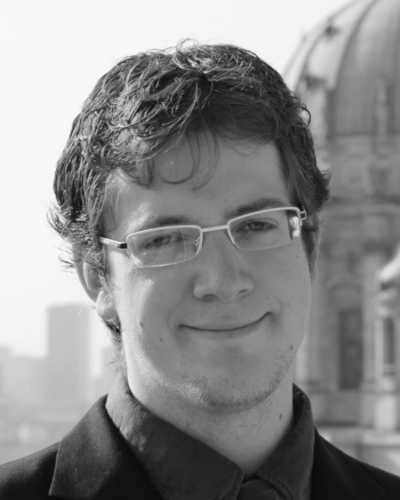}}]{Pierre Guillou}
is a research engineer at Sorbonne Université. After graduating from
MINES ParisTech, a top French engineering school in 2013, he received
his Ph.D., also from MINES ParisTech, in 2016. His Ph.D. work revolved
around parallel image processing algorithms for embedded
accelerators. Since 2019, he has been an active contributor to TTK and
the author of many modules created for the VESTEC project.
\end{IEEEbiography}


\vspace{-1ex}
\begin{IEEEbiography}[{\includegraphics[width=1in,height=1.25in,clip,
keepaspectratio]{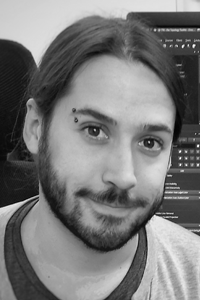}}]{Julien Tierny}
received the Ph.D. degree in Computer Science from the University 
of Lille in 2008 and the Habilitation degree (HDR) from Sorbonne University in 
2016. He is currently a CNRS permanent scientist, affiliated with 
Sorbonne University (Paris, France). Prior to his CNRS tenure, he held a 
Fulbright fellowship (U.S. Department of State) and was a post-doctoral 
researcher at the Scientific Computing and Imaging Institute at the University 
of Utah. 
His research expertise lies in topological methods for data analysis 
and visualization. 
He is the founder and lead developer of the Topology ToolKit 
(TTK), an open source library for topological data analysis.
\end{IEEEbiography}




\end{document}


\renewcommand{\sectionautorefname}{Sec.}
\renewcommand{\subsectionautorefname}{Sec.}
\renewcommand{\figureautorefname}{Fig.}
\renewcommand{\equationautorefname}{Eq.}
\newcommand{\algorithmautorefname}{Alg.}
\renewcommand{\tableautorefname}{Tab.}
\newcommand{\mycaption}[1]{\vspace{-0ex}\caption{#1\vspace{-0ex}}}

\newcommand{\jules}[1]{\textcolor{black}{#1}}
\newcommand{\julien}[1]{\textcolor{black}{#1}}

\newcommand{\julienRevision}[2]{\textcolor{black}{#2}}
\newcommand{\julesRevision}[2]{\textcolor{black}{#2}}

\title{A Progressive Approach to Scalar Field Topology\\Appendix}

\author{Jules~Vidal,
        Pierre~Guillou,
        and~Julien~Tierny
\IEEEcompsocitemizethanks{\IEEEcompsocthanksitem 
J. Vidal, P. Guillou, J. Tierny are with Sorbonne Université and CNRS.
\protect\\
E-mail: \{jules.vidal, pierre.guillou, julien.tierny\}@sorbonne-universite.fr
}
\thanks{Manuscript received April 19, 2005; revised August 26, 2015.}
    }

\markboth{Journal of \LaTeX\ Class Files,~Vol.~14, No.~8, August~2015}%
{Shell \MakeLowercase{\textit{et al.}}: Bare Demo of IEEEtran.cls for Computer Society Journals}

\newcommand{\domain}{\mathcal{M}}
\newcommand{\range}{\mathbb{R}}
\newcommand{\sublevelset}[1]{#1^{-1}_{-\infty}}
\newcommand{\Star}{St}
\newcommand{\Link}{Lk}
\newcommand{\simplex}{\sigma}
\newcommand{\face}{\tau}
\newcommand{\lowerlink}{\Link^{-}}
\newcommand{\upperlink}{\Link^{+}}
\newcommand{\Index}{\mathcal{I}}
\newcommand{\offset}{o}
\newcommand{\Natural}{\mathbb{N}}
\newcommand{\criticalSet}{\mathcal{C}}
\newcommand{\diagram}{\mathcal{D}}
\newcommand{\pointMetric}[1]{d_#1}
\newcommand{\wasserstein}[1]{W_#1}
\newcommand{\projection}{\Delta}
\newcommand{\hierarchy}{\mathcal{H}}
\newcommand{\decimation}{D}
\newcommand{\xDimD}{L_x^\decimation}
\newcommand{\yDimD}{L_y^\decimation}
\newcommand{\zDimD}{L_z^\decimation}
\newcommand{\xDim}{L_x}
\newcommand{\yDim}{L_y}
\newcommand{\zDim}{L_z}
\newcommand{\Grid}{\mathcal{G}}
\newcommand{\GridD}{\mathcal{G}^\decimation}
\newcommand{\x}{\phantom{x}}
\newcommand{\Mod}{\;\mathrm{mod}\;}
\newcommand{\NN}{\mathbb{N}}
\newcommand{\forwardIntegralLine}{\mathcal{L}^+}
\newcommand{\backwardIntegralLine}{\mathcal{L}^-}
\newcommand{\triangulationOp}{\phi}
\newcommand{\decimationOp}{\Pi}



\maketitle

\begin{figure}
  \centering
  \begin{minipage}{0.9\textwidth}
  \centering
  \includegraphics[width=.8\textwidth]{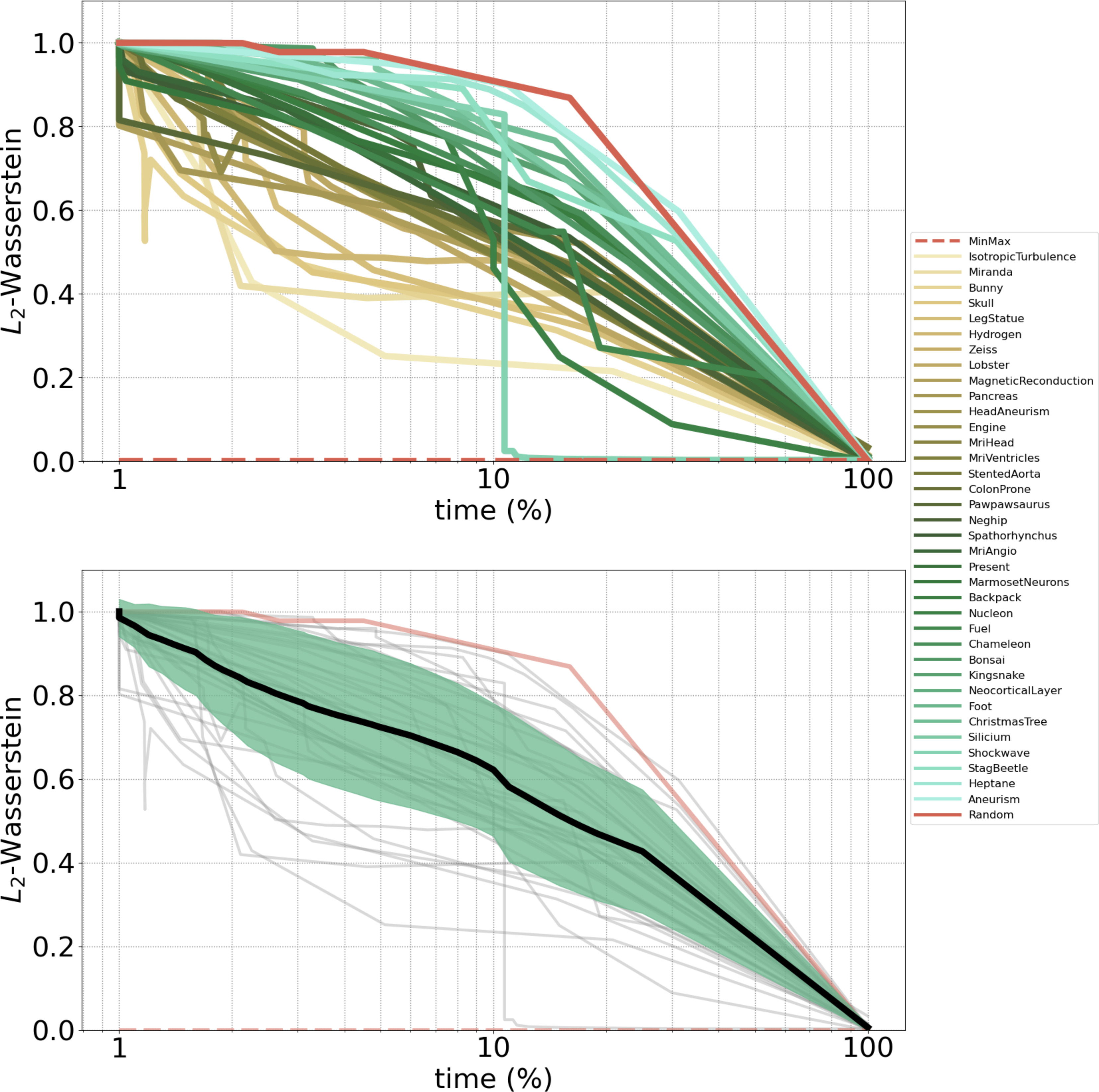}
  \caption{
  \julienRevision{Empirical convergence of the normalized L2-Wasserstein 
distance for an extensive list 
    of data sets.\\}
    {Empirical convergence of the normalized $L_2$-Wasserstein distance for an 
extensive list of datasets.}
\textbf{Top}: each curve plots the distance between the currently estimated 
diagram, 
\julienRevision{D(fi)}{$\diagram(f_i)$}, and the final, exact diagram, 
\julienRevision{D(f)}{$\diagram(f)$}, as 
a function of the percentage of
computation times (logarithmic scale). 
\julienRevision{The color represents the order of curves from lowest average 
distance
(light beige denoting a fast convergence) to highest (light blue, denoting a 
slow convergence).}
{The color map indicates the average distance, from light brown (small 
distances, fast convergence) to light green (large distances, slow 
convergence).}
\julienRevision{The best-case data set (dashed line)
and worst-case data set (plain line) are both plotted in red.\\}
{Extreme synthetic cases are reported in red (dash: \emph{MinMax}, solid: 
\emph{Random}).} 
\textbf{Bottom}: 
\julienRevision{Average curve (black line) of convergence curves for every 
data sets (fine gray lines). The standard deviation around
the mean is represented by the green hull. Stress cases are still in red.}
{Average normalized $L_2$-Wasserstein distance (black curve) and standard 
  deviation (green hull) for all \julesRevision{}{real-life }datasets
  \julesRevision{}{(\textit{i.e. Random} and \textit{MinMax} excluded) }as a function
  of the percentage of 
computation time. Per-dataset curves are shown in the
background (red: synthetic extreme cases, grey: other
datasets).
}
}
\label{fig_openscivis_convergence}
\end{minipage}
\end{figure}

\cleardoublepage

\julienRevision{}
{Figure~\ref{fig_openscivis_convergence} (top) presents a convergence plot 
similar to the one which can be found in the main manuscript, but this time on 
an extensive list of real-life datasets: 
all the datasets from the 
\emph{Open 
Scientific Visualization Dataset}, 
repository,
\href{https://klacansky.com/open-scivis-datasets/}
{https://klacansky.com/open-scivis-datasets/}, 
which fit in the main memory of 
our experimental setup. This represents a set of $36$ datasets (containing 
acquired and simulated data). 
As discussed in the paper, while 
the 
monotonic decrease of the $L_2$-Wasserstein distance along the computation 
cannot be guaranteed at a theoretical level, it can still be observed in 
practice.
Note however that for two examples (\emph{Bunny} and 
\emph{Engine}) a slight oscillation can be observed in the early stages of the 
computation (between $1\%$ and $2\%$ of the computation time). However, past 
this 
point, the distance keeps on decreasing monotonically. Also, note that the 
convergence curves for all datasets are indeed located between the 
curves of the two extreme synthetic examples (\emph{MinMax} and \emph{Random}). 
The bottom part of Figure~\ref{fig_openscivis_convergence}, which reports the 
average distance for 
all datasets and its standard deviation, further 
confirms the overall convergence tendency.}
